\documentclass[useAMS, usenatbib]{mn2e}
\usepackage{xspace}
\usepackage{amsmath}
\usepackage{epsfig}
\usepackage{defs}

\def\ie{{\it i.e.}}
\def\eg{{\it e.g.}}
\def\ltsima{$\; \buildrel < \over \sim \;$}
\def\simlt{\lower.5ex\hbox{\ltsima}}
\def\gtsima{$\; \buildrel > \over \sim \;$}
\def\simgt{\lower.5ex\hbox{\gtsima}}

\def\hide#1{}

\title[Clues on the Missing Sources of Reionization]{Clues on the Missing Sources of Reionization from Self-consistent Modeling of Milky Way and Dwarf Galaxy Globular Clusters}

\author[H. Katz \& M. Ricotti]{Harley Katz$^{1,2,3}$\thanks{E-mail: hk380@ast.cam.ac.uk} and Massimo Ricotti$^{1,4}$\thanks{E-mail: ricotti@astro.umd.edu}\\
$^1$Department of Astronomy, University of Maryland, College Park, MD 20742, USA\\
$^2$Institute of Astronomy, University of Cambridge, Madingly Road, Cambridge, CB3 0HA, UK\\
$^3$Kavli Institute for Cosmology, University of Cambridge, Madingly Road, Cambridge, CB3 0HA, UK\\
$^4$Sorbonne Universités, Institut Lagrange de Paris (ILP), 98 bis Bouldevard Arago 75014 Paris, France\\
} 

\begin{document}

\maketitle

\begin{abstract}
Globular clusters are unique tracers of ancient star formation.  We
determine the formation efficiencies of globular clusters across
cosmic time by modeling the formation and dynamical evolution of the globular cluster population of a
Milky Way type galaxy in hierarchical cosmology, using the merger tree
from the Via Lactea II simulation. All of the models are constrained
to reproduce the observed specific frequency and initial mass function
of globular clusters in isolated dwarfs. Globular cluster orbits are
then computed in a time varying gravitational potential after they are
either accreted from a satellite halo or formed {\it in situ}, within
the Milky Way halo.

We find that the Galactocentric distances and metallicity distribution
of globular clusters are very sensitive to the formation efficiencies
of globular clusters as a function of redshift and halo mass.  Our
most accurate models reveal two distinct peaks in the globular cluster
formation efficiency at $z \sim 2$ and $z\sim 7-12$ and prefer a
formation efficiency that is mildly increasing with decreasing halo mass, the
opposite of what expected for feedback-regulated star formation.  This
model accurately reproduces the positions, velocities, mass function,
metallicity distribution, and age distribution of globular clusters in
the Milky Way and predicts that $\sim 40\%$ formed {\it in situ},
within the Milky Way halo, while the other $\sim 60\%$ were accreted
from about 20 satellite dwarf galaxies with $v_{cir}>30$~km/s, and
about $29$\% or all globular clusters formed at redshifts $z>7$.

These results further strengthen the notion that globular cluster
formation was an important mode of star formation in high-redshift
galaxies and likely played a significant role in the reionization of the
intergalactic medium.
\end{abstract}


\section{Introduction}

Globular Clusters (GCs), often referred to as fossils of ancient star
formation, are compact gravitationally bound systems of $\sim 10^6$
stars which typically orbit a much larger host galaxy.  These systems
are among the oldest bound stellar objects known to exist, with some
forming only a few hundred million years after the big bang.  While
these systems have been very well studied in the local Universe, their
formation and evolutionary histories remain open questions in modern
astrophysics.

Nearly all GCs are homogeneous in heavy elements \citep{Sneden2005},
which suggests that the majority of their stars formed in an
instantaneous burst with a high efficiency of gas-to-star conversion
\citep{James2004, Carretta2009a}.  Despite the homogeneity of heavy
elements within individual GCs, these systems are often classified
into two categories: metal-poor with [Fe/H] $\leq-1.5$ and metal rich
with [Fe/H] $\geq -1.5$.  The origin of these two populations is
unknown; however, this bimodality has been observed in
multiple galactic environments \citep{Zepf1993}.  Furthermore, the
ages and kinematics of GCs also exhibit characteristic bimodal
distributions.  

Past observations have shown that galactic metal-poor GCs tend to be older than
metal-rich GCs and the age spread in metal-poor GCs is $\sim 1$~Gyr
compared to the $\sim 6$~Gyr dispersion in metal-rich GCs
\citep{Rosenberg1999, Salaris2002, Marin2009}.  There is some evidence
of self enrichment in GCs; however, the age gap between metal-poor and
metal-rich GCs is greater than the range present within each
population suggesting that these are two distinct populations
\citep{Marin2009}.  However, more recent observations have shown a gradual trend of increasing age spread with increasing metallicity which weakens this age gap described in the previous works \citep{Dotter2011,VBerg2013}.  Two different branches in the age versus metallicity plane have been identified which maintains the notion of two distinct populations.

Kinematically, multiple observations reveal that
the red, metal-rich GCs are more spatially concentrated than the blue,
metal-poor GC population \citep{Pota2013, Schuberth2010, Faifer2011,
  Strader2011}.  Additionally, the metal-poor GCs tend to rotate less
than the metal-rich GC population and the rotation of the metal-rich
GCs has been associated with the photometric axis of the host galaxy
\citep{Pota2013}.  The distinct characteristics of these two
populations suggests that they likely formed at different epochs and
under different conditions.
 
Due to their old stellar populations, GCs today are faint and
difficult to detect at large distances. However, \cite{Katz2013} have
shown that for $5-10$~Myrs after their formation, GC systems are very
bright and can be detected in deep fields even at redshift
$z=8$. Although these systems are spatially unresolved, their UV
luminosity functions and UV continuum slopes can be modeled and have
characteristic signatures due to their bursting mode of star
formation. It is therefore possible to set meaningful upper limits on
their formation rate across multiple redshifts from $z=8$ to
$z=1$. \cite{Katz2013} concluded that GCs likely formed in two
distinct epochs: $z>6$ and $z<3$.  Additionally, Monte Carlo
simulations convolving the age estimates of the Milky Way GCs with
Gaussian $\pm 1$~Gyr uncertainties, support the notion of a bimodal
formation history.  \cite{Katz2013}'s results and the observed bimodal
properties of these systems, suggest that there existed two distinct
epochs of GC formation, with the old population possibly important for
reionizing the intergalactic medium \citep{Ricotti2002}. However, it
is unclear if this scenario is consistent with the observed properties
of Milky Way GCs and observations of GC systems in nearby isolated
dwarfs. How did the Milky Way come to posses its current population of
GCs? What fraction formed {\it{in situ}} within the Milky Way and what
fraction was accreted onto the Milky Way via tidal disruption of
merging dwarf galaxies?  What role could GCs have played in the reionization of the Universe?

Many simulations have already been run attempting to address these questions.  \cite{PG08} populated an N-Body simulation of a Milky Way type galaxy in order to determine if the ages, masses, metallicities and kinematics could be reproduced.  The simulations were very successful in reproducing many of the observations however the model could not put good constraints on the when the GCs form and the mean distances of GCs were farther out than what is observed.  \cite{Griffen2010} modeled both metal poor and metal rich GC formation in the Aquarius simulations by identifying both likely sites of GC formation in haloes with $T>10^4$ K as well as a population that formed in the mergers of haloes.  The formation epochs of GCs were constrained based on the role that the already formed GCs played in the reionization of the local medium.

We also aim to answer these questions by modeling GC formation in a high
resolution N-body simulation of a Milky Way type galaxy and comparing
characteristics of the resulting GC population with those properties
exhibited by the Milky Way's GCs.  Tuning the few free parameters in
our simulations reveal new and unexpected insights into galaxy
formation.  We take a different approach than some of the previous simulations by matching the properties of the Milky Way GC population simultaneously to the characteristics of isolated dwarf GCs, which allows us
to constrain the formation efficiencies of GCs across cosmic time and
thus determine the role GCs may have played in the evolution of the
Universe. This method is completely independent of the one used in
\cite{Katz2013}, but interestingly seems to point to similar results.

All observational data for GCs was compiled using catalogs from the
following references:
\cite{Geo2009a,Geo2009,Forbes2010,Gnedin1997,Strader2011,Galleti2004,Peacock2010,Harris1996}
(2010) Edition.  In addition, this work made use of catalogs from
\cite{VizieR}.

The paper is organized as follows: In Section~2, we describe the
mechanisms responsible for the dynamical evolution of GCs.  In
Section~3, we propose a shape for the GC Initial Mass Function (GCIMF)
as well as provide analytical calculations involving observations of
GCs in local dwarfs to constrain the minimum formation efficiencies of
GCs.  We describe our simulations in Section~4 and interpret their
results in Section~5.  In Section~6, we present our
discussion.  In Section~7 we calculate the contribution of GCs to reionization and in Section~8, we discuss our conclusions.

\section{Mechanisms for the Dynamical Evolution of Globular Clusters}

The GC population that we observe today is likely a poor
representation of the original GC population of a galaxy.  Stellar
evolution and dynamical effects, including two-body relaxation,
dynamical friction, tidal shocks, and tidal truncation significantly
reduce the number and mean mass of GCs from their epoch of formation
to the present \citep{Ostriker1997,Fall2001}.  Our treatment of the
dynamical evolution of GCs follows closely the one in \cite{PG08},
with a few differences that will be emphasized as we describe the
details of the model.

We assume that all dynamical effects are independent of each other and
the GC's mass, $M_{gc}$, is governed by the following differential
equation:
\begin{equation}
\frac{dM_{gc}}{dt}=-(\nu_{se}(t)+\nu_{ev}(t)+\nu_{sh}(t))M_{gc},
\end{equation}
where $\nu_{se}$, $\nu_{ev}$, $\nu_{sh}$ are the respective mass-loss
rates due to stellar evolution, two-body relaxation, and tidal shocks
\citep{Fall2001,PG08}.  While the assumption that these processes are
independent is simplistic, it is certainly well motivated due to the
distinct time scales over which each mechanism operates.

We adopt a Kroupa stellar IMF for the GCs \citep{Kroupa2001} which has
a mean mass of $\bar{m}\approx 0.387$~M$_{\odot}$ after the high mass
stars ($M>2$~M$_{\odot}$) have died off.  Within the first $300$~Myr,
about 30\% of the initial mass of the GC is lost due to stellar
evolution. This time scale is very short compared to the typical ages
of GCs.  For a detailed description and calculation of stellar
evolution, we refer to section~3.1 of \cite{PG08}.

For two-body relaxation, we refer to the approximation derived by \cite{Spitzer1987},
\begin{equation}
\nu_{ev}=\frac{7.25\xi_e\bar{m}G^{1/2}\ln\Lambda_{cl}}{M_{gc}^{1/2}R_h^{3/2}},
\label{eq:nu_ev}
\end{equation}
where $\xi_e=0.045$ is a normalization factor derived by
\cite{Henon1961}. The parameter $\ln \Lambda_{cl}$ is the
time-dependent coulomb logarithm which is derived by \cite{Binney1987}
to be between 10 and 12 and we have chosen $\ln \Lambda_{cl}=12$.  As
described later, our simulated GCs are modeled with constant density
as they evolve, which causes the mass loss due to two-body relaxation
alone to be constant as a function of time for all GCs (i.e. $\nu_{ev}
\propto M_{gc}^{-1}$).  Furthermore, this process is ineffective at
destroying GCs during the first few hundred Myrs in the GC life cycle
and only becomes important at later times.  Because the time scales
over which stellar evolution and two-body relaxation operate are
clearly distinct, we treat them independently.

For disk shocking we refer to the approximation in \cite{PG08} with a slight modification:
\begin{equation}
\nu_{sh}=
\begin{cases}
\frac{5/3}{\Delta t}\frac{I_{tid}R_h^3}{GM_{gc}}\ &{\text if}\ sign(z(n-1)) \neq sign(z(n)),\\
0\ &{\text if}\ sign(z(n-1)) = sign(z(n)),
\end{cases}
\end{equation}
where $n$ is the current time step corresponding to a time $t$ and $n-1$ is the time corresponding to the previous time step.  This ensures that disk shocking is only effective when the sign of the z-coordinate changes between time steps indicating that the GC has crossed the plane of the disk.  $\Delta t$ is the length of the time step in our integration
routine for the GCs orbits and $I_{tid} \sim 4g_m^2/V_z^2$.
The parameter $g_m$ is the maximum vertical acceleration and $V_z$ is
the component of the GC's velocity orthogonal to the disk.  It is clear that this is an instantaneous process and can therefore be
treated separately from both stellar evolution and two-body
relaxation.

When the half mass radius, $R_h$, of a GC approaches its tidal radius,
$R_t$, large percentages of stars can be stripped, significantly
decreasing its mass.  The tidal radius of the GC then shrinks due to
this loss in mass, allowing more stars to escape.  This process is
unstable, resulting in the destruction of the GC on a relatively short
timescale \citep{Baumgardt1998}.  \cite{Baumgardt1998} define a
critical value $x_{crit}$ such that if the ratio $R_h/R_t \geq
x_{crit}$ then the GC is destroyed.  We choose $x_{crit}=0.37$, which
is more conservative than what was chosen in \cite{Baumgardt1998}.  This was tuned to correspond to less than a few hundred pc from the galactic center in our model and we believe this assumption is well motivated given that no GCs are observed within a few hundred parsecs of the galactic center.  Furthermore, we emphasize that regardless of the choice of $x_{crit}$, since we will model the GCs as having constant density, this process effects all GCs in our model equally. As we will later discuss, this mechanism only controls the overall normalization of GCs in our model which is a parameter that we can only bound with upper and lower limits rather than make an explicit prediction.
We define $R_t$, consistent with \cite{Baumgardt1998} as:
\begin{equation}
R_t \equiv \left(\frac{M_{gc}}{3M_{gal}(<R_{gc})}\right)^{1/3}R_{gc},
\end{equation}
where $R_{gc}$ is the distance of the GC from the galaxy and $M_{gal}$
is the mass of the galaxy inside $R_{gc}$.  

Finally, we also compute the acceleration due to dynamical friction
using the Chandrasekhar formula \citep{Chandra1943} as given by
\cite{Binney1987} assuming that the $M_{gc}>>m$, where $m$ is the mass
of the dark matter particles (when the gravitation potential is
dominated by the dark matter halo) or bulge stars:
\begin{equation}
\frac{d\vec{v}_{gc}}{dt}=-\frac{4\pi G^2\rho M_{gc}\ln\Lambda}{v_{gc}^3}\left[erf(X)-\frac{2X}{\sqrt{\pi}}e^{-X^2}\right]\vec{v}_{gc},
\end{equation}
where here we define $\Lambda \equiv {b_{max}V_C^2}/(GM_{gc})$ with
$b_{max}$ being the impact parameter, and $V_C$ is the speed of the GC
through the galaxy which we calculate by integrating the orbit, but
can also be roughly estimated as the local circular velocity. Here,
$\rho$ is the local density of the galaxy, $X \equiv
v_{gc}/(\sqrt{2}\sigma)$ and $\sigma$ is the velocity dispersion of
the dark matter particles (or bulge stars).

\begin{figure*}
\centerline{\epsfig{figure=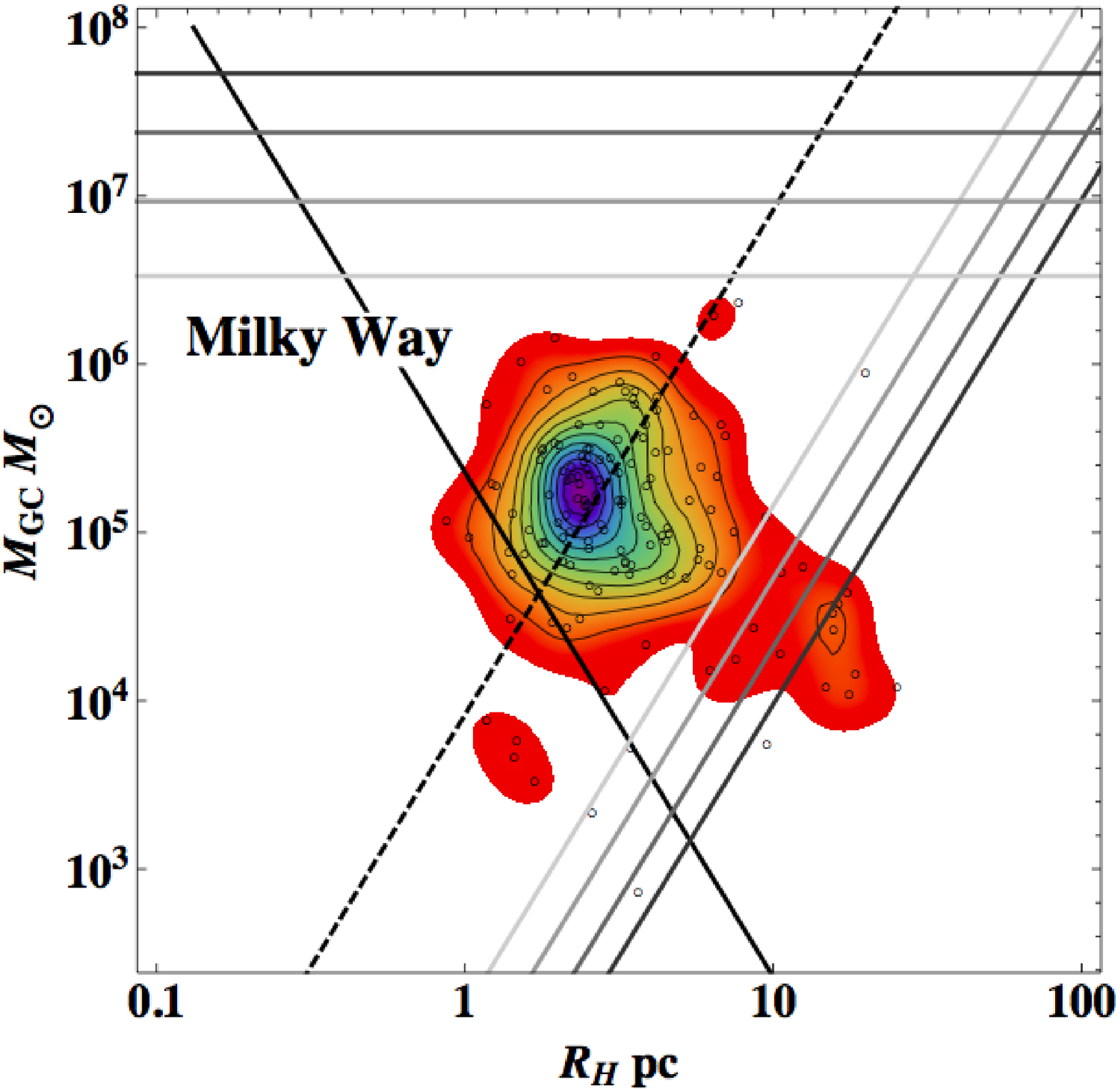, width=6cm}\epsfig{figure=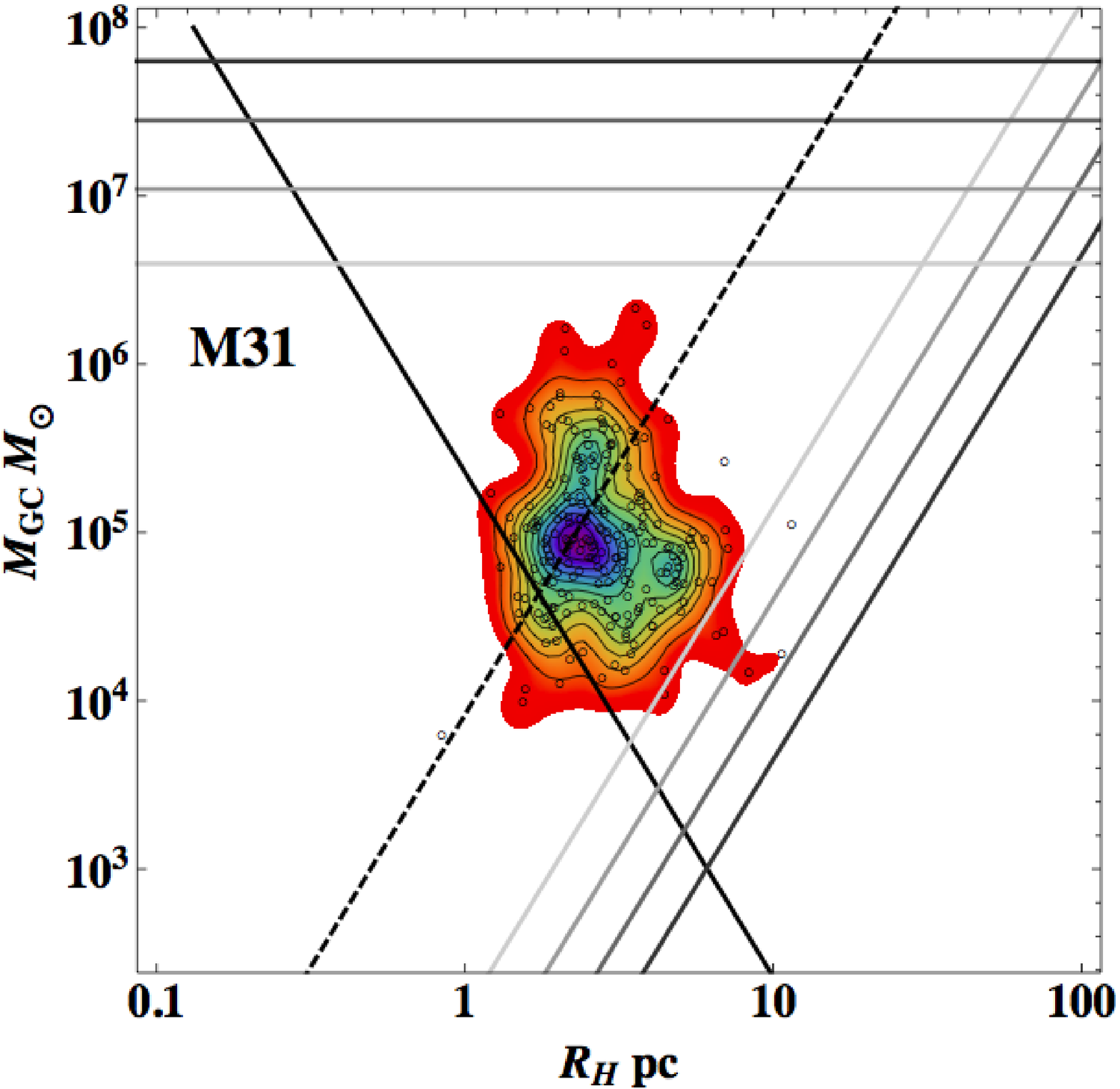, width=6cm}\epsfig{figure=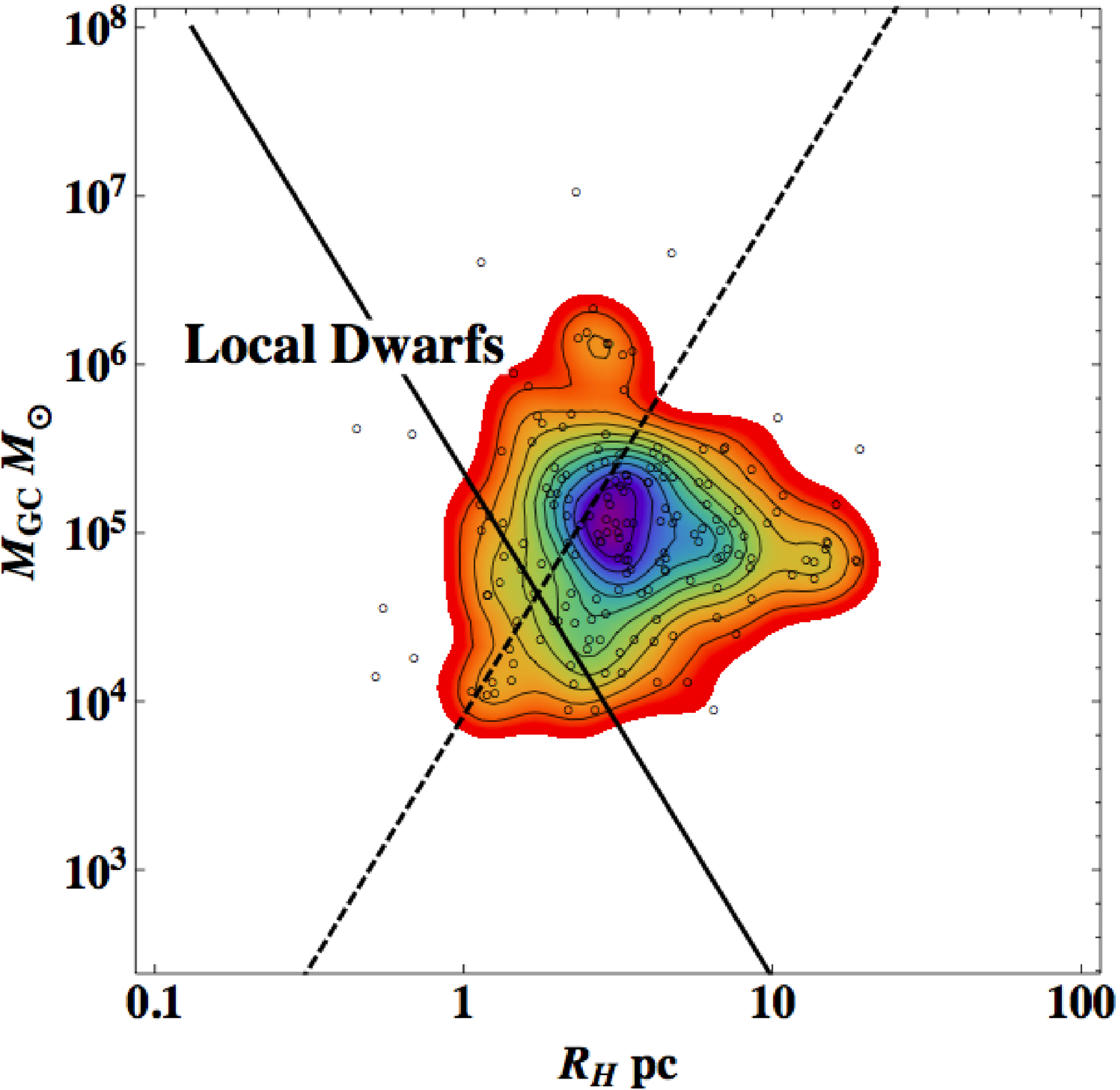, width=6cm}}
\caption{{\it Left.} Vital diagram for Milky Way GCs. {\it Center.}
  Vital diagram for M31. {\it Right.} Vital diagram for GCs in local
  dwarf galaxies.  For the data we show both the data point and the
  normalized density distribution function. The left side of the
  triangle is the theoretical line for relaxation.  The right side
  lines are the theoretical lines for tidal destruction and the top
  lines are theoretical lines for dynamical friction.  Different lines
  for tidal destruction and dynamical friction represent different
  initial positions for the GCs corresponding to 12, 7, 5, 3 kpc
  respectively.  All of these lines were computed with the assumption
  that the GCs are 12~Gyr old.  The majority of Milky Way's and M31's
  GCs fit in this triangle which verifies our assumptions are valid.
  We do not plot the top and right sides of the triangle for the GCs
  in local dwarfs because these lines are unique for each galaxy
  whereas our sample is from multiple different dwarfs.  The dotted
  black line in the left and center panels represent the constant
  density we use to model GCs in our simulations. Data for the Milky
  GCs is from \protect\cite{Geo2009a}, data for M31 is from
  \protect\cite{Strader2011} and data for GCs in local dwarf galaxies
  is from \protect\cite{Geo2009}.  The mass model we use for the Milky
  way is from \protect\cite{Irrgang2013}.}
\label{vitaldiag}
\end{figure*}

For an approximate test of the validity of these approximations, we
can take an analytical approach to modeling the destruction of 140
Milky Way GCs using their measured masses and half light radii.
Plotted in Figure~\ref{vitaldiag} is the ``vital diagram", first
created by \cite{Ostriker1997} for both the Milky Way and M31.  Since
almost all GCs reside within the triangle defined by lines of constant
dynamical friction timescale (upper horizontal line), two-body
relaxation timescale (left side of the triangle), and tidal
destruction timescale (right side of the triangle), we can be
confident that our approximations are fairly accurate.

\subsection{Dynamical Evolution in Dwarf Galaxies}\label{sec:dyn}

\begin{figure}
\epsfig{figure=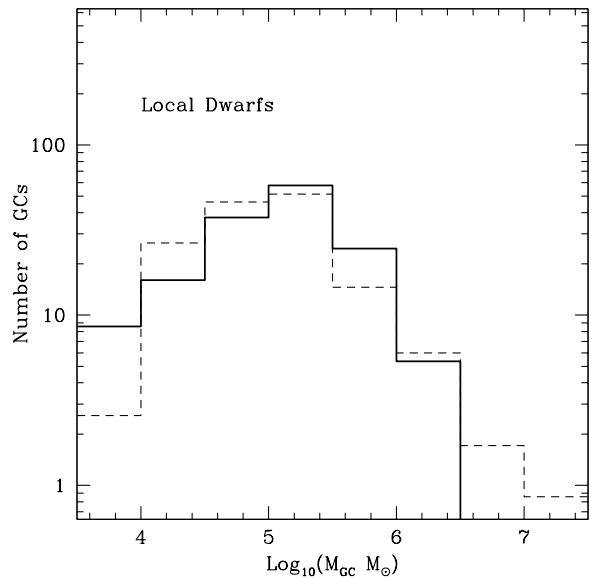,scale=0.4}
\caption{Mass function of GCs in local dwarf galaxies from
  \protect\cite{Geo2010} (dashed line) compared to that of the Milky
  Way (thick line).  The histogram has been normalized to a population
  of 150 GCs.  The excess of low mass GCs may suggest that tidal
  effects are inefficient in disrupting GCs in dwarf galaxies.}
\label{massfunc_dwarf}
\end{figure}

While these approximations are accurate for the Milky Way and M31 GC
populations, one might ask whether the same can be said for GCs in
dwarf galaxies.  The naive expectation is that dynamical friction
should be enhanced while tidal destruction is minimized.  It is
interesting to note that the GCs in dwarf galaxies tend to occupy
roughly the same locations in $M-R_h$ space as the GCs in the Milky
Way and M31, as shown in the right panel of Figure~\ref{vitaldiag}.
It seems there is a slight tendency for GCs in the local dwarfs to
occupy some of the lower density space compared to GCs in M31 and the
Milky Way, pointing to a less efficient tidal destruction rate due to
the lower density of dwarf galaxies.  Furthermore, we note that the
mass function of GCs in dwarf galaxies shows a higher abundance of GCs
at lower mass compared to the Milky Way, as illustrated in
Figure~\ref{massfunc_dwarf}.  It is these GCs which are more
susceptible to destruction from tides and this higher abundance may
also point to tides being inefficient in dwarf galaxies.

The Fornax dwarf spheroidal galaxy is the Milky Way's largest dwarf
spheroidal and it has a system of five GCs ranging in mass from
$3.7\times10^4~M_{\odot}-3.63\times10^5~M_{\odot}$ at distances of
$0.24-1.6$~kpc \citep{Angus2009}.  Multiple studies \citep{Goerdt2006,
  Read2006, Sanchez2006, Inoue2009} have claimed that these GCs should
have fallen to the center of the host galaxy in much less than a
Hubble time due to dynamical friction.  The Fornax dwarf spheroidal
does not show evidence of a bright nucleus where the sinked GCs would
reside \citep{Angus2009}. We found similar contradictory results on
the effect of dynamical friction in dwarfs looking at a large sample
of dwarf galaxies from \cite{Geo2010}.

Multiple groups, such as the ones listed previously have attempted to
resolve this issue but there has not been an agreed upon conclusion.
While this outcome clearly demonstrates a lack of understanding of the
system, it provides an interesting prospect: if dynamical friction is
ineffective, GCs survive in their host galaxy longer than expected and
are more easily accreted onto a larger galaxy if the dwarf galaxy
falls into a deeper gravitational well.  In addition, if GCs reside
closer to the tidal radius of the dwarf galaxy, then as soon as the
dwarf approaches a much larger galaxy, the GCs will be accreted nearly
instantly.  Although we do not understand why the GCs are not sinking
in the Fornax dwarf spheroidal and other local dwarfs, this seems a
common pattern among GCs in dwarf galaxies. Hence, a large fraction of
old metal-poor GCs in the Milky Way have likely been accreted from
dwarf satellites rather than formed {\it{in situ}}.

\section{Constrain the GC's Mass Function and Formation Efficiencies from Dwarf Galaxies}

\subsection{The GC's Mass Function in Dwarf Galaxies}
\label{ssec:mf}

The exact shape and normalization of the GCIMF remains reasonably
unconstrained; however, multiple groups have shown that a
power law or Gaussian GCIMF can lead to the right shape for the GCs in
the Milky Way when the dynamical evolution is simulated
\citep{Fall2001,PG08}. Dwarf galaxies offer an ideal environment to
determine the initial mass function and formation efficiency of GCs.
Some dynamical processes that are important in more massive galaxies,
like the Milky Way, can be neglected, therefore rendering the
calculations more robust. Since the effects of tidal shocking and
tidal truncation are minimized for high density GCs in isolated dwarf
galaxies, the high mass end of the GC mass function in dwarf galaxies
should be representative of the GCIMF.

Modeling the evolution of the GC mass function in dwarf galaxies is
thus quite simple. Stellar evolution destroys $\sim 30\%$ of a GC mass
within the first few Myr of its lifetime, and using our assumption
that each GC maintains its initial density during its evolution,
two-body relaxation is easily modeled as a constant mass loss
proportional to the age of the GC.  If we suppose that the shape of
the GCIMF is a power law with slope $\alpha$, the subsequent evolution
transforms the mass function to the form:
\begin{equation}
\frac{dN}{dM}\propto (M+M_{2bd})^\alpha,
\label{eq:2bd}
\end{equation}
where $M_{2bd} \propto t$ is the mass loss due to two-body
relaxation. For $M \gg M_{2bd}$, this equation reduces to $dN/dM
\propto M^\alpha$ and for $M \ll M_{2bd}$, we get $dN/dM \propto
M_{2bd}^\alpha = const$.

\begin{figure*}
\centerline{\epsfig{figure=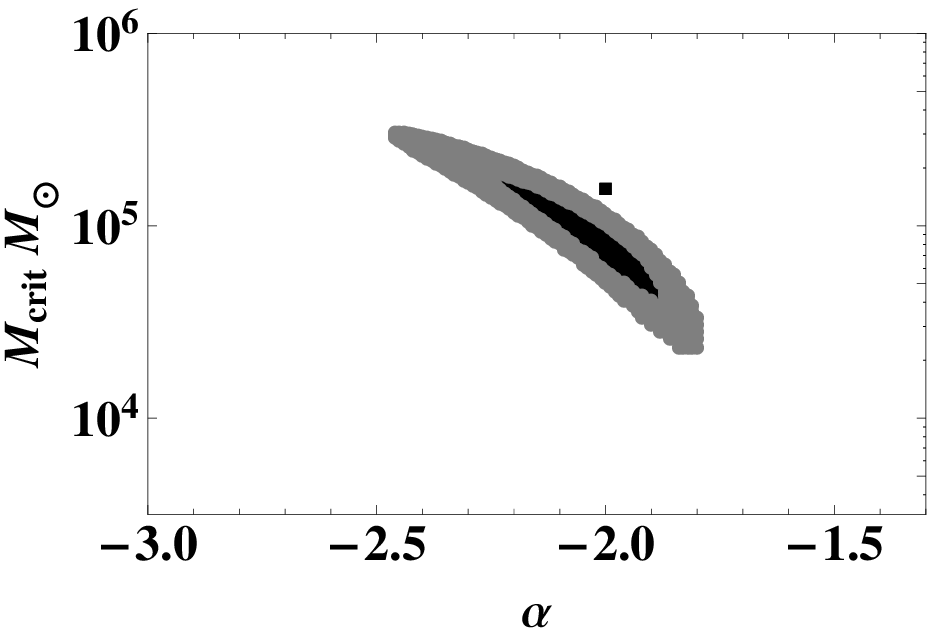,scale=0.87}\epsfig{figure=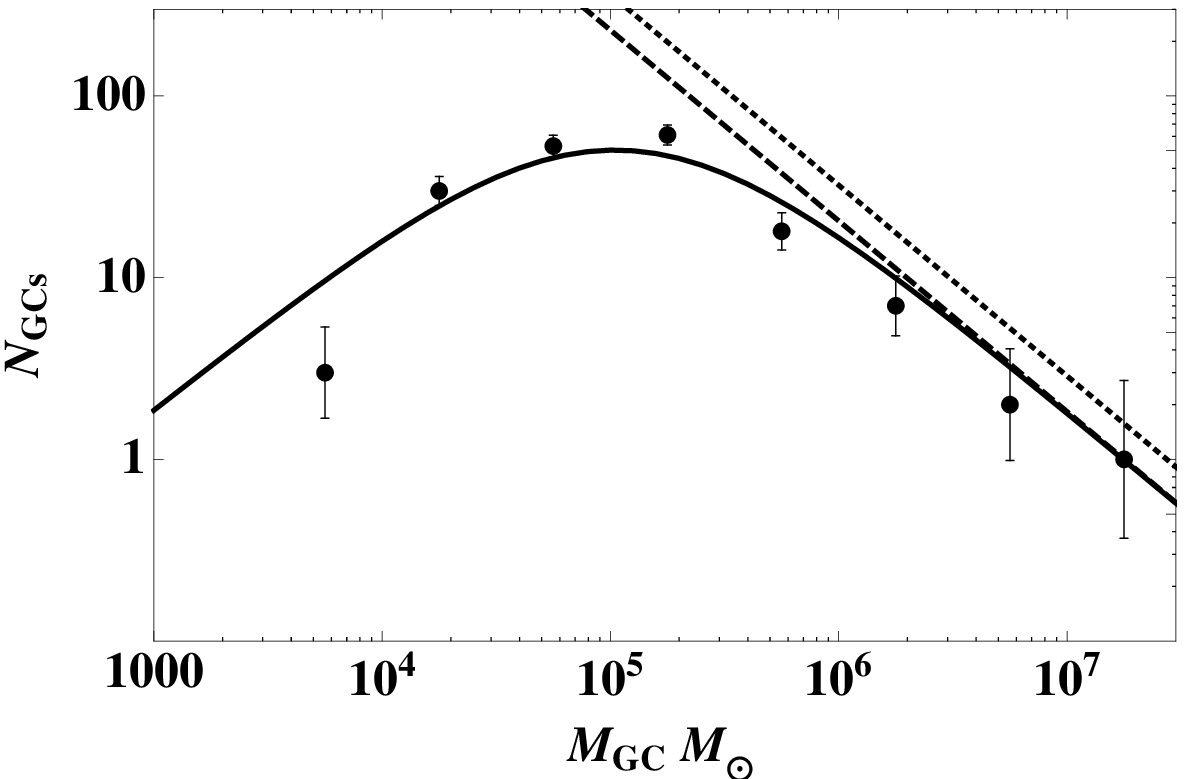,scale=0.7}}
\caption{{\it Left.} Parameter space analysis of the best fit slope
  and critical mass for the GCIMF for the constant density model.  The
  black and gray regions are the $1\sigma$ and $2\sigma$ confidence
  levels respectively. The square represents the parameters used in
  \protect\cite{PG08}. {\it Right.}  Our chosen GCIMF parameters
  compared with the dwarf galaxy GCMF for the constant density model.}
\label{minx2}
\end{figure*}

Letting the slope of the GCIMF and the mass loss due to two-body
relaxation be free parameters, we can constrain the density of GCs as
well as the slope of the GCIMF by fitting the mass function of GCs in
local dwarf galaxies.  In the left panel of Figure~\ref{minx2}, we
show the one and two sigma confidence limits for $M_{2bd}$ and $\alpha$.
The best fit values are $\alpha=-2.05$ and
$M_{2bd}=1.1\times10^5\ M_{\odot}$. Assuming and average age of GCs of
$12$~Gyrs, using equation~(\ref{eq:nu_ev}) we derive an average GC density of
$2000$~M$_{\odot}$/pc$^3$, which is consistent with the average
density of GCs in both the Milky Way and M31 (see
Figure~\ref{vitaldiag}).  In the right panel of Figure~\ref{minx2} we
show the best fit model for the GC's mass function compared data from
local dwarf galaxies.

In Appendix~A we relax the assumption that all GCs have constant mean
density.  We keep the assumption that GCs maintain a constant density
as they evolve, but we explore the case in which the initial density
of each GC is related to their mass.  To test this idea, we assume
that all GCs have a constant $R_h$ so that $\rho_h \propto M_{gc}$.
However, using this model we are unable to match simultaneously the
shape of the mass function in dwarfs and the Milky Way GCs. For this
reason we only consider the constant density model in the rest of this
paper.

\subsection{Constraints on GC's Formation Efficiencies}
\label{subsec:constraints}

While the shape of the GCIMF dictates the relative efficiency of low
mass and high mass GC formation, the overall normalization remains
unconstrained.  The bottom panel of Figure~4 in \cite{Geo2010} shows
that late type and early type dwarf galaxies in their sample with absolute visual
magnitude brighter than $-16.5$ contain at least one observed GC.
This data, in combination with the mass loss rates and
destruction rates of GCs, estimated in the previous section, can be
used to constrain the GC formation efficiency as a function of
redshift for both red and blue dwarf galaxies.

Stellar evolution alone removes about 30\% of the initial mass of all
GCs within the first hundred few Myr, assuming a Kroupa stellar
IMF. Thus, given an initial total mass in GCs, $M_{gc}^{ini}$, the
effect of stellar evolution is to keep the number of GCs constant but
reduce the total mass in GCs and each individual GC mass by a factor of
$0.7$. Assuming a GCIMF with a power-law slope of $\alpha=2$ as
derived in \S~\ref{ssec:mf}, the mean GC mass is $\langle
m_{gc}\rangle^{ini}=M_{low}\ln(M_{up}/M_{low})$, and thus, the initial
number of GCs is $N^{ini}=M_{gc}^{ini}/\langle
m_{gc}\rangle^{ini}$. For our fiducial model with
$M_{low}=10^5$~M$_\odot$ and $M_{up}=2.857\times10^7$~M$_\odot$, we
have $\langle m_{gc}\rangle^{ini} = 5.2 \times 10^5$~M$_\odot$.
 
The number of GCs that survive two-body relaxation and mass loss due to
stellar evolution can be estimated analytically from $dN/dM$ given in
equation~(\ref{eq:2bd}):
\begin{equation}
\begin{split}
N^{surv}=\int^{M_{up}}_{M_{low}^*} \frac{dN}{dM} dM\\
=N^{ini}\frac{M_{low}}{M_{low}^*}=N^{ini}Min[1,0.7 M_{low}/M_{2bd}],
\end{split}
\end{equation}
where we define $M_{low}^*=Max[M_{low},M_{2bd}/0.7]$.  We define the
number of surviving GCs, $N^{surv} \equiv f_N^{surv} N^{ini}$, for our
fiducial model and $M_{2bd}=1.1 \times 10^5(t/12~Gyr)$~M$_\odot$
derived in Section~\ref{ssec:mf} and we, therefore, find
\begin{equation}
f_N^{surv}(t)=Min\left[1, 63.6\%\left(\frac{M_{low}}{10^5M_{\odot}}\right)\left(\frac{12~Gyr}{t}\right)\right].
\end{equation}
Similarly, the surviving fraction by mass is
\begin{equation}
\begin{split}
f_M^{surv}(t)&=\frac{\langle m_{gc}\rangle^{2bd}}{\langle m_{gc}\rangle^{ini}} f_N^{surv}\\
& \approx \frac{56\%}{1+\ln(10^5~M_\odot/M_{low})/5.65} ~\text{at t=12~Gyr},
\end{split}
\label{eq:fM}
\end{equation}
where
\begin{equation}
\begin{split}
& \langle m_{gc}\rangle^{2bd} = 0.7 M_{low}^*\ln\left[\frac{M_{up}}{0.7 M_{low}^*}\right]-M_{2bd}\\
& =
\begin{cases}
0.7\langle m_{gc}\rangle^{ini}-M_{2bd} & ~\text{for $M_{2bd}<0.7M_{low}$,}\\
M_{2bd}\left(\ln\left[\frac{M_{up}}{M_{2bd}}\right]-1\right) & ~\text{for  $M_{2bd}>0.7M_{low}$.}
\end{cases}
\end{split}
\end{equation}
For the fiducial values adopted here $\langle m_{gc}\rangle^{2bd}=5
\times 10^5$~M$_\odot$ at $t=12$~Gyr.  Our calculation assumes that
tidal effects in dwarf galaxies can be neglected and the mass loss due
to stellar evolution and two-body relaxation alone represents the
majority of the mass loss.  Assuming, as explained later, that the
total mass in GCs at formation\footnote{We have normalized this
  equation to a mass of $7\times 10^9$~M$_{\odot}$ because this is the
  normalization we use in our simulations which corresponds to a $V_C$
  of 50 km/s.} is
\begin{equation}
M_{gc}^{ini} \equiv \eta_i 7 \times 10^9~M_\odot \left(\frac{M_{dm}}{7\times10^9~M_\odot}\right)^{1+\gamma},
\label{eqn:formed}
\end{equation}
and adopting the analogous definition for the formation efficiency $\eta$ after evolutionary effects we have
\begin{equation}
\eta = \eta_i f_M^{surv}.
\label{eq:eta}
\end{equation}
Finally, adopting a mass-to-light relationship for dwarfs,
$L_V\approx2.6\times10^8~L_\odot (M_{dm}/7\times10^9~M_\odot)^{5/3}$, as in \cite{Geo2010}, we obtain for the fiducial model:
\begin{equation}
N_{gc}^{surv}(t)= f_N^{surv} \eta_i \left(\frac{7\times10^9}{\langle m_{gc} \rangle ^{ini}}\right)\left(\frac{L_V}{2.6\times10^7~L_\odot}\right)^{\frac{3(1+\gamma)}{5}}.
\label{eq:numb}
\end{equation}

From Figure 4 in \cite{Geo2010}, it is clear that red dwarf galaxies
have a larger $N_{gc}^{surv}$ than the blue ones at fixed luminosity,
for $M_V\leq-12$. Within the working assumption that the adopted
mass-to-light ratio of the dwarfs is the same for blue and red
galaxies, this result points to a larger $\eta_i$ for early type dwarf
galaxies (also consistent with assuming a larger $\eta_i$ at high
redshift). Additionally, the number of surviving GCs remains
relatively constant with decreasing luminosity for early type dwarfs,
pointing to a small value of $1+\gamma$, or $-1< \gamma < 0$.

Inspecting Figure~4 in \cite{Geo2010} we see that all (red and blue)
dwarf galaxies with luminosity $L \sim 5.4\times 10^8$~L$_\odot$ have
at least one GC. At this luminosity we set $N_{gc}^{surv} = 1$ to
derive a rough estimate of the minimum formation efficiency to produce
at least one surviving GC. The slope of the curve at $M_V\leq-12$ for
the fraction of dwarfs with at least one observed GC as a function of
luminosity gives $\gamma \sim -0.8$. There are only a few galaxies in
the magnitude bins dimmer than $M_V=-12$ and thus we only fit this
curve for brighter magnitude bins.  From Equation~(\ref{eq:numb}) we
derive the efficiency of GC formation in early type dwarfs to be
\begin{equation}
\eta_i(t)^{red}\approx 6.8\times10^{-5}f_N^{surv}(t)^{-1}.
\end{equation}
 Similarly, for late type dwarfs we obtain $\gamma \sim -0.35$ and
\begin{equation}
\eta_i(t)^{blue} \approx 5.5\times 10^{-5}f_N^{surv}(t)^{-1}.
\end{equation}
In conclusion, these two values of $\eta_i$ and $\gamma$ bracket the
mean formation efficiency of GCs in dwarf galaxies as a function of
time. These results also point to a {\it GC formation efficiency
  either constant or increasing with decreasing galaxy mass}, the
opposite of what expected due to feedback effects for ``normal'' star
formation in dwarf galaxies. In addition, the large scatter (of nearly
two orders of magnitude) of the specific frequency at a fixed galaxy
luminosity may be accommodated in models in which $\eta_i$ varies as a
function of cosmic time by a similar amount.  In Figure~\ref{etaevo},
we plot how minimum values of $\eta_i$ vary with time.

Note that in \cite{Geo2010} and other studies, the GC formation
efficiency, $\eta$, is defined as $M_{gc}^{surv} \equiv \eta M_{dm}$,
analogously to our definition of $\eta$ given by
Equation~(\ref{eq:eta}) assuming $\gamma=0$.  \cite{Geo2010} have
found that $\eta \approx 5 \times10^{-5}$ which is consistent with
\cite{Blakeslee1999}, who found $\eta \approx 1.71 \times10^{-4}$, and
other studies including, \cite{Kravtsov2005} and \cite{Spilter2009}.
This value is the observed efficiency which is post stellar evolution
and dynamical processes.  The initial GC formation efficiency prior
to stellar evolution and dynamical processes, $\eta_i$, is a free
parameter in our simulations. For whatever $\eta_i$ we choose, we are
constrained by observations of dwarf galaxies to a range of values
$10^{-5}<\eta<10^{-4}$.

\begin{figure}
\epsfig{figure=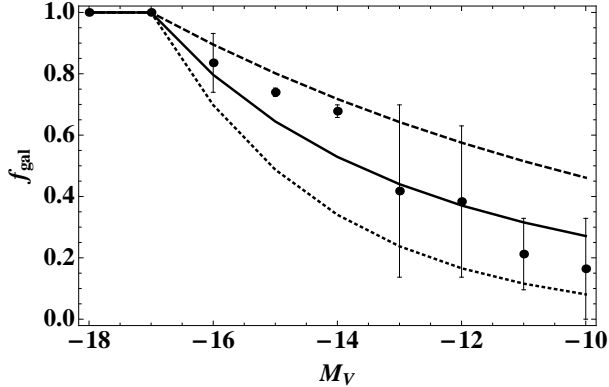,scale=0.65}
\caption{The percentage of galaxies expected to host at least one GC
  as a function of the absolute visual magnitude of the host galaxy
  for 12 Gyr old GCs.  The dashed line represents the expectations for
  red galaxies and the dotted line represents the expectations for
  blue galaxies.  The black line is the mean of the two types of
  galaxies.  The data points are the average of the blue and red
  galaxies that host at least one GC from \protect\cite{Geo2010}.
  Error bars on data points correspond to the range between red and
  blue galaxies and the true dispersion in the mean of all galaxies is
  likely larger than what is plotted.}
\label{fgal}
\end{figure}

\begin{figure}
\epsfig{figure=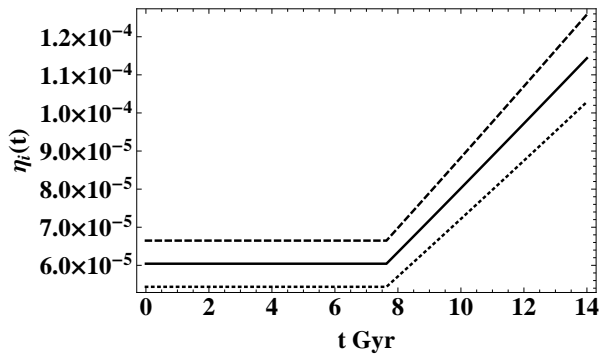,scale=0.65}
\caption{We show the minimum formation efficiencies, $\eta_i(t)$, as a
  function of time for our best fit GCIMF for the UDM.  The dashed
  line is for red galaxies, the dotted line is for blue galaxies, and
  the black line is the mean of the two types of galaxies.}
\label{etaevo}
\end{figure}

\section{Modeling GC Formation, Accretion, and Dynamical Evolution}

\subsection{The Via Lactea II Simulation}

We use the halo merger tree from a high resolution N-body simulation,
the Via~Lactea~II (VL~II) \citep{Diemand2008}, to model the formation,
accretion, and dynamical evolution of GCs in a Milky Way type halo and
in a few isolated dwarfs around the main halo. The VL~II simulation is
a cosmological N-body simulation of over 1 billion particles with mass
resolution of $4.1 \times 10^3$~M$_{\odot}$. The simulation starts at
$z=104.3$ in a 40 comoving Mpc periodic box using cosmological
parameters from WMAP~3 \citep{Spergel2007}.  At $z=0$, the main halo
is representative of a Milky Way type galaxy with $r_{200}=402.1$~kpc,
$M_{200}=1.917 \times 10^{12}$~M$_{\odot}$ and a maximum circular
velocity of $V_{cir,max}=201.3$ km/s (note that here the subscripts
refer to mean halo density $200$ times the cosmic value, not $200$
times the critical density).  We utilize the publicly available
evolutionary tracks of the 20,000 largest haloes and subhaloes at
$z=4.56$ as well as the complete merger tree (Diemand private
communication).  The time evolution is sampled uniformly
after $z=7.77$, but rather coarsely ($\Delta t = 687$~Myr) with 27
outputs from the simulation beginning at $z=27.54$ to the end at
$z=0$.

\subsection{Distributing GCs to Haloes}

As described in Section~\ref{subsec:constraints} and in
Equation~(\ref{eqn:formed}), we choose to parameterize the number (and
total mass) of GCs in a halo, at any given time, as a fraction of the
dark matter mass of the host. The number of GCs we attribute to each
halo at the time of virialization (which we define to be when a halo reaches its maximum $V_{max}$) is:
\begin{equation}
N_{GC}=\tilde \eta_i(z,M_h)\frac{M_h}{\langle{m}_{gc}\rangle^{ini}},
\label{eq:ngc}
\end{equation}
with specific GC formation efficiency
\begin{equation}
\tilde \eta_i(z,M_h) \equiv \eta_i(z)\left(\frac{M_h}{7 \times 10^9~M_\odot}\right)^\gamma,
\end{equation}
where $\eta_i(z)$ is a function of redshift, and $\gamma$ is a free
parameter describing the dependence of the GCs formation efficiency on
the halo mass. We choose a pivot point for the dependence of $\eta_i$
on halo mass appropriate for dwarf galaxies.  Because the masses of
haloes in the simulation are environmentally dependent (and somewhat
arbitrarily defined in N-body simulations), we choose to relate
$N_{GC}$ to the maximum circular velocity of the halo, $V_{max}$,
rather than the halo mass. This assumption is also convenient to
define a redshift of formation of a halo and whether a subhalo is
being tidally stripped. Indeed $V_{max}$ reaches its maximum value at
the redshift of virialization and remains roughly constant afterwards,
unless the halo is being tidally stripped.  Hence, we define $M_h$
consistent with the parameters for the Milky Way as follows:
\begin{equation}
M_h=1.25\times10^{12}~M_{\odot}\left(\frac{V_{max}}{220\ km/s}\right)^{\beta^\prime},
\label{eq:mh}
\end{equation}
where $\beta^\prime$ in general depends on the halo mass (and redshift
of formation) and has a value between 3 and 4. We adopt a fiducial
value $\beta^\prime=3.5$ that gives $M_h=7 \times
10^9~M_\odot(V_{max}/50~km/s)^{3.5}$, also appropriate for dwarf
galaxies. Combining Equations~(\ref{eq:ngc})-(\ref{eq:mh}), we define
our model for {\it in situ} formation of GCs:
\begin{equation}
N_{GC}=\eta_i(z)\frac{7\times10^{9}~M_{\odot}}{\langle{m}_{gc}\rangle^{ini}}\left(\frac{V_{cir,max}}{50\ km/s}\right)^{\beta},
\label{oldeqn}
\end{equation}
where we allow for $\beta \equiv \beta^\prime (1+\gamma) \not=
\beta^\prime$, if the GC formation efficiency depends on the halo mass
in addition to the redshift of virialization of the host halo. Values
$\beta < \beta^\prime$ indicate that dwarf galaxies have a higher {\it
  in situ} GC formation efficiency per unit halo mass than Milky Way
sized galaxies. We will explore this model in
Section~\ref{subsec:aspen2}.

Equation~(\ref{oldeqn}) also determines the minimum circular velocity
of a halo that forms at least one GC.  Because $\eta_i$ varies with
redshift, the minimum circular velocity defined by this equation is a
function of $z$ and $\beta$.  We have chosen $\beta=3.5$ (\ie,
$\gamma=0$) for our fiducial model. However, as discussed in
Section~\ref{subsec:constraints}, data on local dwarf galaxies by
\cite{Geo2010} show that early type and late type dwarfs have
different values of $\gamma$ and $\eta_i$, hence suggesting that these parameters depend on redshift.

We assign GCs to the haloes when they have reached their maximum
$V_{max}$, which indicates that they have just virialized. Other
groups, including \cite{PG08} define a truncation redshift, $z_t$,
after which GCs can no longer form.  This is certainly a major assumption of our model as the formation mechanisms of GCs have yet to be resolved (excluding the GCs we see forming in mergers which are discussed later).  The formation epochs we define for GCs should have little effect on the kinematics of the final surviving population because we will demonstrate that all GCs in our models are equally susceptible to tidal effects.  As long as the GCs are formed prior to dwarf being accreted onto the main halo, our results will be robust to this parameter.  Small variations in the formation epochs will only have marginal effects on the mass function because two body relaxation is the only mechanism for which the time dependence is explicit, prior to the time of accretion onto the main halo.  We refrain from advocating for one of the many proposed mechanisms for the formation of GCs as this is not the focus of the current work.  We emphasize that as long as the GCs have formed in the dwarf haloes prior to the accretion epoch, our model is very robust.

We assign a mass to each GC by randomly drawing from the chosen GCIMF.
We adopt a constant density model for the internal structure of each
GC.  While this model may seem too simplistic, \cite{PG08} found that
their model using GCs with a constant half-mass density of
$\rho_{GC}=4\times10^3$~M$_{\odot}$pc$^{-3}$ best reproduced the mass
function of the Milky Way metal poor GCs. In Section~\ref{ssec:mf} we found
similar results testing a constant and a variable initial density
model.  Here we also adopt the constant density model however
with a slightly different density $\rho_{GC}\approx2
\times10^3$~M$_{\odot}$pc$^{-3}$ constrained to reproduce the mass
function of GCs in local isolated dwarfs (see Section~\ref{ssec:mf}).

\subsection{Accretion and Dynamical Evolution}

We divide the GCs in our simulation into three distinct categories:
1) GCs which form in a specific halo that survive to the present
without merging with a larger halo (\ie, {\it in situ} formation of GCs in
dwarf galaxies or in the main halo). 2) GCs formed in haloes which are
accreted by another halo and remain outside the virial radius of
the Milky Way halo (\ie, GCs accreted by dwarf galaxies). 3) GCs which are
eventually accreted by the Milky Way halo.
\begin{enumerate}
\item For the first class of GCs, we compute stellar evolution and
  two-body relaxation beginning at the epoch of formation.  A GC is
  considered destroyed when its mass becomes less than
  $10^4$~M$_{\odot}$.  This mass threshold is chosen by looking at the
  catalog of GCs in nearby dwarf galaxies \citep{Geo2010}.  There are
  very few GCs with mass less than $10^4$~M$_{\odot}$, which is likely
  a result of the rate of destruction being fast at these masses and
  partially an observational bias. We do not compute the GCs' orbits
  in this case because GCs are weakly affected by tidal destruction in
  low surface brightness dwarf galaxies.  The densities of the GCs are
  larger than the stellar and dark matter densities in dwarfs and
  justifying why minimal tidal destruction is to be expected for this
  specific population of GCs. Thus stellar evolution and two-body
  relaxation are the main modes of GC destruction in dwarfs, allowing
  us to analytically reproduce the mass function of GCs in nearby
  dwarf galaxies.

  The one exception to this assumption holds for the
  GCs that form {\it in situ} in the Milky Way's progenitor halo, or
  along with the main halo in the simulation.  For these GCs, we
  compute the orbits as well as dynamical friction, tidal shocks, and
  tidal destruction along with two-body relaxation and stellar
  evolution as described in Section~\ref{sec:dyn}.  For the GCs that
  form {\it in situ} in the Milky Way, we distribute them randomly in
  a uniform density sphere within 1.5 disk scale lengths and with
  velocities smaller than the local $V_C$ so that they are dynamically
  cold.  We have tested the effect of changing the volume of the
  sphere and the magnitude of the initial velocity and note that there
  is not much difference except when the GCs are placed well within the
  bulge or much farther than the disk size.  We find our simulations
  to be largely insensitive to these parameters.

\item The second class of GCs is very similar to the first, but
  includes GCs accreted onto dwarf galaxies from satellites, in
  addition to {\it in situ} formation. For the same reasons previously
  stated, we do not compute orbits and evolve the GC mass function
  with only stellar evolution and two-body relaxation.  We define the
  accretion epoch to be the redshift at which $V_{max}$ of the
  satellite halo becomes smaller than a fraction $f=70\%$ of the
  maximum $V_{max}^{max}$ over all redshifts.  At this time, tidal
  interactions are affecting the inner part of the halo and we assume
  that GCs are stripped from their host and accreted.  In order to
  determine the halo onto which the stripped GCs are accreted, we use
  the full merger tree of the simulation.  We search for the closest
  halo with a mass greater than our tidally disrupted halo and define
  this as the new host. GCs can undergo multiple accretion events
  throughout cosmic time.

\item The third class of GCs is perhaps the most interesting, as these
  are GCs that are eventually accreted onto the Milky Way halo.  Prior
  to accretion, we compute stellar evolution and two-body relaxation
  as in the previous two categories.  However, if a halo's $V_{max}$
  drops to below $f=70\%$ of its maximum $V_{max}^{max}$ and this halo
  is within the virial radius of the Milky Way halo, the GCs are
  accreted onto the Milky Way halo and we begin computing their orbits
  including the effect of dynamical friction, tidal destruction, and
  tidal shocks.  We use the velocity of the disrupted host as the
  initial velocity for all accreted GCs associated with this halo.
  Additionally, we offset the position of each GC in the satellite
  halo by randomly assigning their position within spherical shells.
  We limit the radial coordinates of each GC to be within the tidal
  radius at the time output before accretion and the tidal radius at
  the epoch of accretion.  It is important to note that our
  simulations results are rather insensitive to our choice of
  $f$.  This is likely because the orbits of accreted GCs depend
  mainly on the velocity of the GCs host at the time of the accretion
  rather than its exact position along the orbit. The position of the
  GCs when accreted are not well resolved due to the coarse time
  resolution of our merger tree, that consists of only 27 redshift
  outputs, that we use to interpolate the orbits of galaxies and
  satellites.  This coarse time step is unlikely to fully
  resolve the infall of the satellite due to dynamical friction,
  possibly leading to a slight error on the position and time in which
  we define the destruction epoch for each dwarf halo.

We note that our definition of accretion epoch is slightly different than some other uses in the literature.  \cite{PG08} begin orbits of GCs at the point when the dwarf enters within the virial radius of the main halo.  \cite{Griffen2010} consider the GCs as part of the main halo their simulations when the GC are within twice the half mass radius of the main halo at z=0.  The epoch of accretion of the dwarf would certainly be when the dwarf becomes gravitationally bound to the main halo; however, the important definition for this work is the epoch at which the globular clusters are stripped from the dwarf haloes (which is what we have defined as the accretion epoch).  This is more likely to be at the time when the dwarf begins to be tidally disrupted as we have described.
\end{enumerate}

We use a time evolving model for the simulated Milky Way galaxy that
includes a dark matter halo, a disk, and a bulge.  The evolution of the dark
matter halo is taken directly from the 27 outputs of the VL~II
simulation.  We use cubic splines to interpolate between the
simulation outputs.  We use a Miyamoto-Nagai potential \citep{MN1975}
to describe the disk and a Plummer sphere to describe the bulge \citep{Plummer1911}:
\begin{equation}
\Phi_{halo}(r)=-\frac{GM_*}{r_s}\frac{ln(1+(r/r_s))}{r/r_s},
\end{equation}
where $M_*=4\pi\rho_s r_s^3$ and $\rho_s=4\rho(r_s)$,
\begin{equation}
\Phi_{disk}(R,z)=-\frac{GM_{disk}}{\sqrt{R^2+(a_d+\sqrt{z^2+b_d^2})^2}},
\end{equation}
\begin{equation}
\Phi_{bulge}(r)=-\frac{GM_{bulge}}{\sqrt{r^2+b_b^2}}.
\end{equation}
We adopt a scaling relation between the mass of the disk to be $4.5\%$
of the mass of the halo at all redshifts, slightly less than
\cite{Klypin2002,PG08}, and supported by observations.  Thus, the mass
of the disk at a given redshift is $M_d(z)=(0.045M_h(z))$.  We choose
$a_d(z)=0.01r_{200}(z)$ and $b_d(z)=0.054a_d(z)$, consistent with the
values for the Milky Way \citep{Pac1990} and a smaller bulge than \citep{Pac1990}  which is also scaled with time.  We note that it is more likely that the Milky Way has a bar rather than a classical bulge and since we use a classical bulge, we satisfy the constraint that the mass of the bulge is not greater than $\sim8\%$ of the mass of the disk and use an extended bulge \citep{Shen2010}.  Because the mass of the bulge is low compared to the rest of the galaxy, only orbits that approach the center are likely effected; however, these GCs are also more prone to destruction given the higher density of the bulge compared to the rest of the galaxy.  In summary, the $z=0$ Milky
Way parameters we adopt are: $M_{200}=1.94\times10^{12}$~M$_{\odot}$,
$r_{200}=462.274$~kpc, $M_{disk}=8.60\times10^{10}$~M$_{\odot}$,
$a_d=4.623$~kpc, $b_d=0.25$~kpc, $M_{bulge}=5\times10^{9}$~M$_{\odot}$ and $b_b=0.540$~kpc.

We use a leap frog integrating scheme with a constant time step of
1~Myr (determined by convergence tests) to follow the orbits of the
GCs around the time evolving potential. We compute stellar evolution,
two-body relaxation, dynamical friction, tidal shocks, and tidal
truncation concurrently for each GC.  We assume that a GC is destroyed
when its mass drops below $M_{gc}=10^4$~M$_{\odot}$, because of the
observed scarcity of Milky Way's GCs with masses below this threshold.

\section{Simulation Results}

\begin{figure}
\epsfig{figure=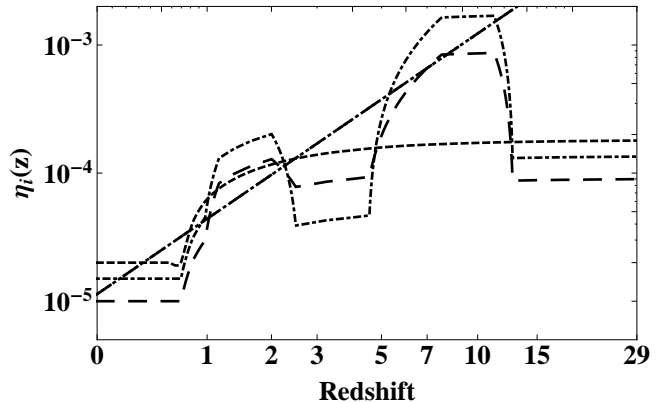,scale=0.7}
\caption{Formation efficiencies for the CE (short dashed), KR13 (long
  dashed), KR13-bis (short dot dashed), and Power Law (long dot
  dashed) models.}
\label{formeff}
\end{figure}

In this section we compare the results of four different models for
the star formation efficiency $\eta_i(z)$ as a function of redshift
and halo mass (parameter $\gamma$). The assumed $\eta_i(z)$ as a
function of redshift are shown in Figure~\ref{formeff}.

\begin{figure*}
\centerline{\epsfig{figure=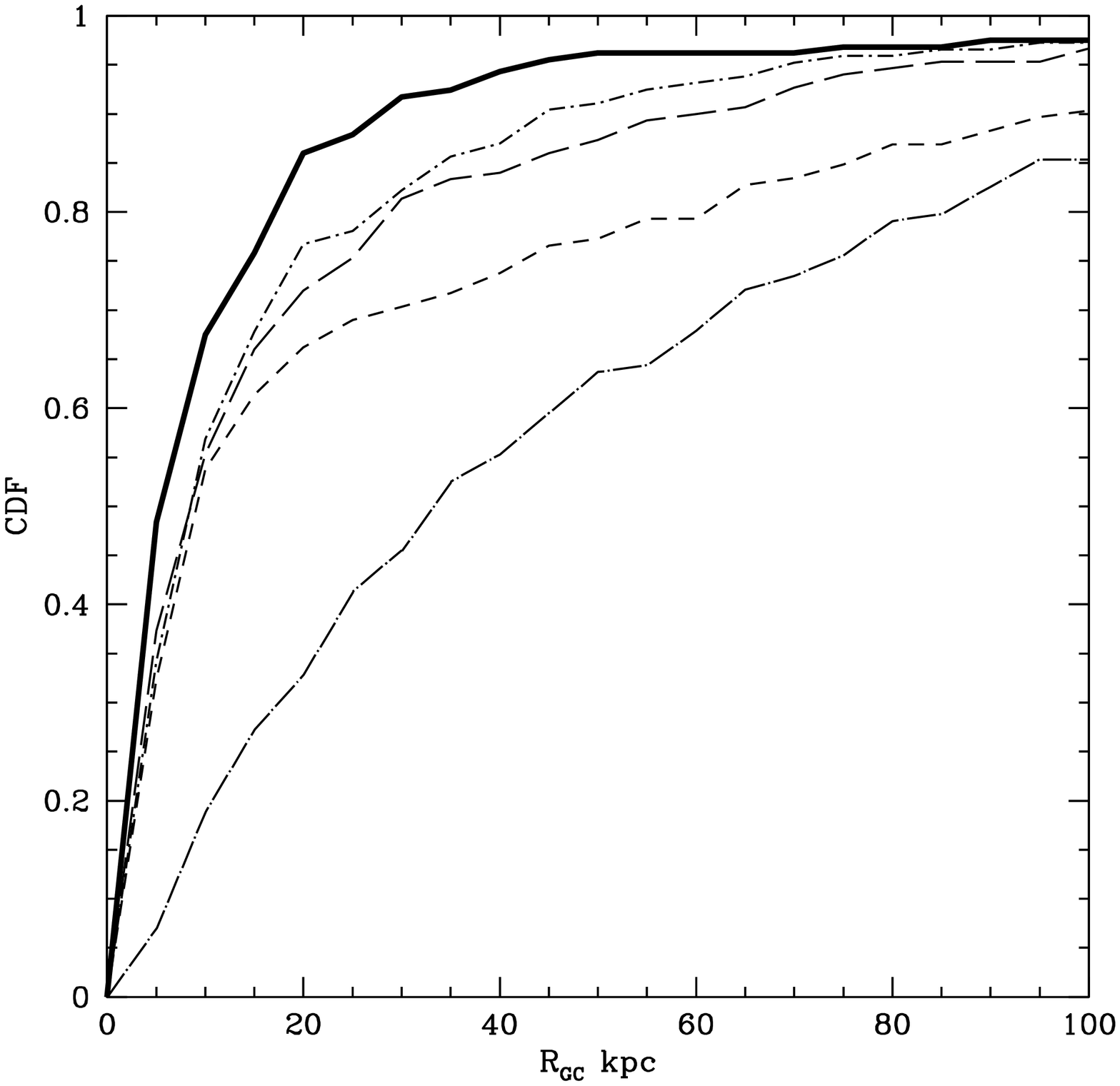,scale=0.45}\epsfig{figure=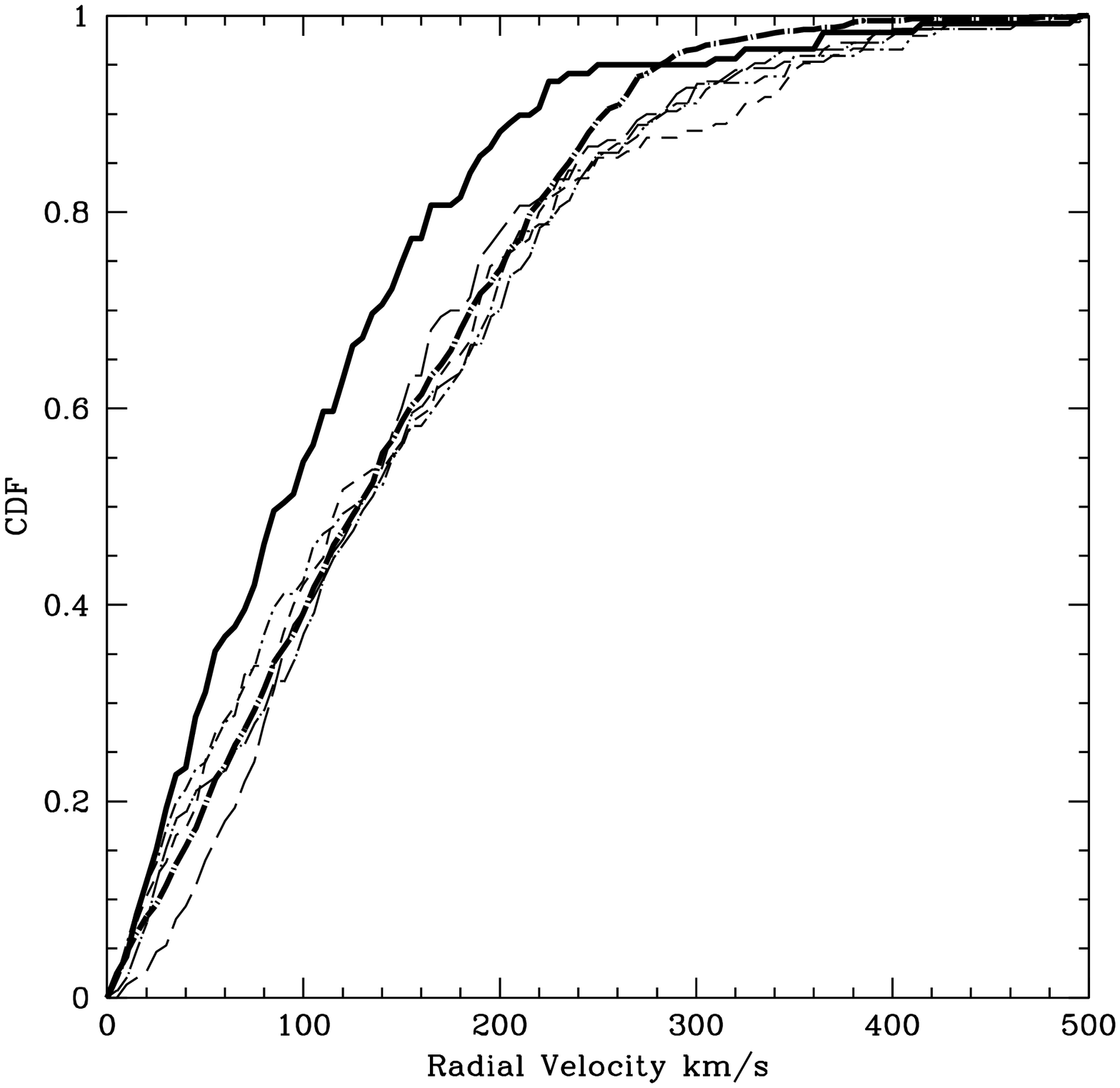,scale=0.45}}
\caption{{\it Left.} The CDF of the Galactocentric distance for GCs in
  the Milky Way (solid), in our CE model (short dashed), in the KR13
  model (long dashed), in the KR13-bis model (dot dashed).  {\it
    Right.} The CDF of the radial velocities of GCs in the Milky Way
  (solid), in M31 (dot dashed), in our CE model (short dashed), in the
  KR13 model (long dashed), in the KR13-bis model (short dot dashed),
  in the Power Law model (long dot dashed).  The positions of GCs in
  our model are much more sensitive to the model than the
  velocities. }
\label{CDFs}
\end{figure*}

\subsection{Constant Formation Efficiency Model}

In our first model we keep the formation efficiency $\eta_i(z)$ nearly
constant as a function of redshift and halo mass (\ie, $\gamma=0$,
$\beta=3.5$), and we determine its value by reproducing the observed
number of GCs in the Milky Way. We denote this model as the Constant
Formation Efficiency model (CE).

In Figure~\ref{CDFs} (left), we plot the cumulative distribution
function (CDF) of Galactocentric distances of the GCs produced by the
CE model.  In this simulation, we form
145 surviving GCs, in good agreement with observations (\ie, $\sim
150$), but this model produces more GCs at large Galactocentric
distances than what is observed in the Milky Way (see the left panel of Figure \ref{CDFs}).  
In addition to the Galactocentric positions,
the velocity distribution of the GCs should not drastically differ
from that of the Milky Way GCs or that of a similar spiral galaxy.  In
Figure~\ref{CDFs} (right), we plot the CDF of the radial velocities of
GCs in our simulation compared with the CDF of the radial velocities
of the Milky Way and M31 GC populations and note a close agreement
between our simulation and the Milky Way.

\begin{figure*}
\epsfig{figure=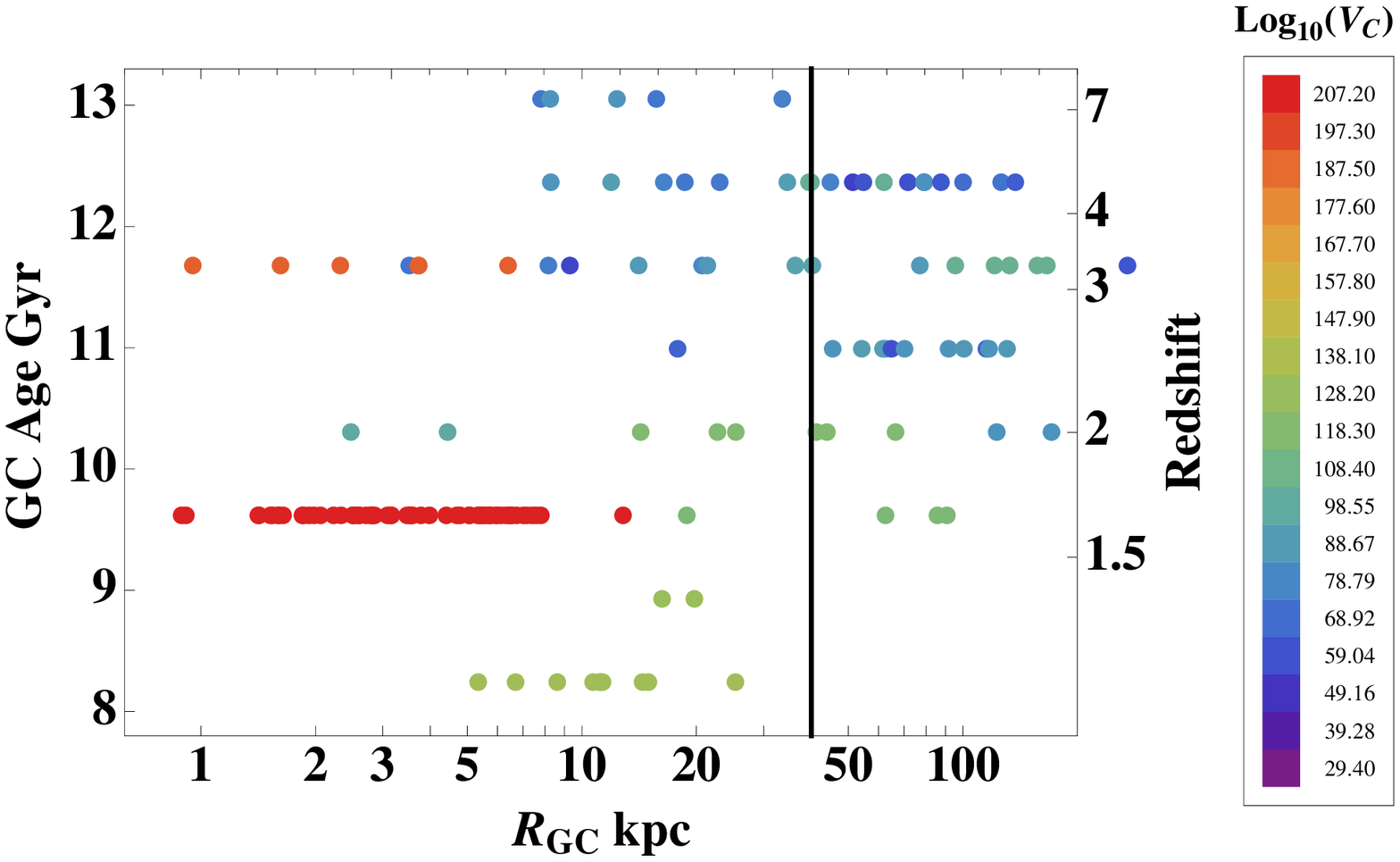,scale=0.42,trim= 280 0 50 0}\ \ \ \ \ \ \epsfig{figure=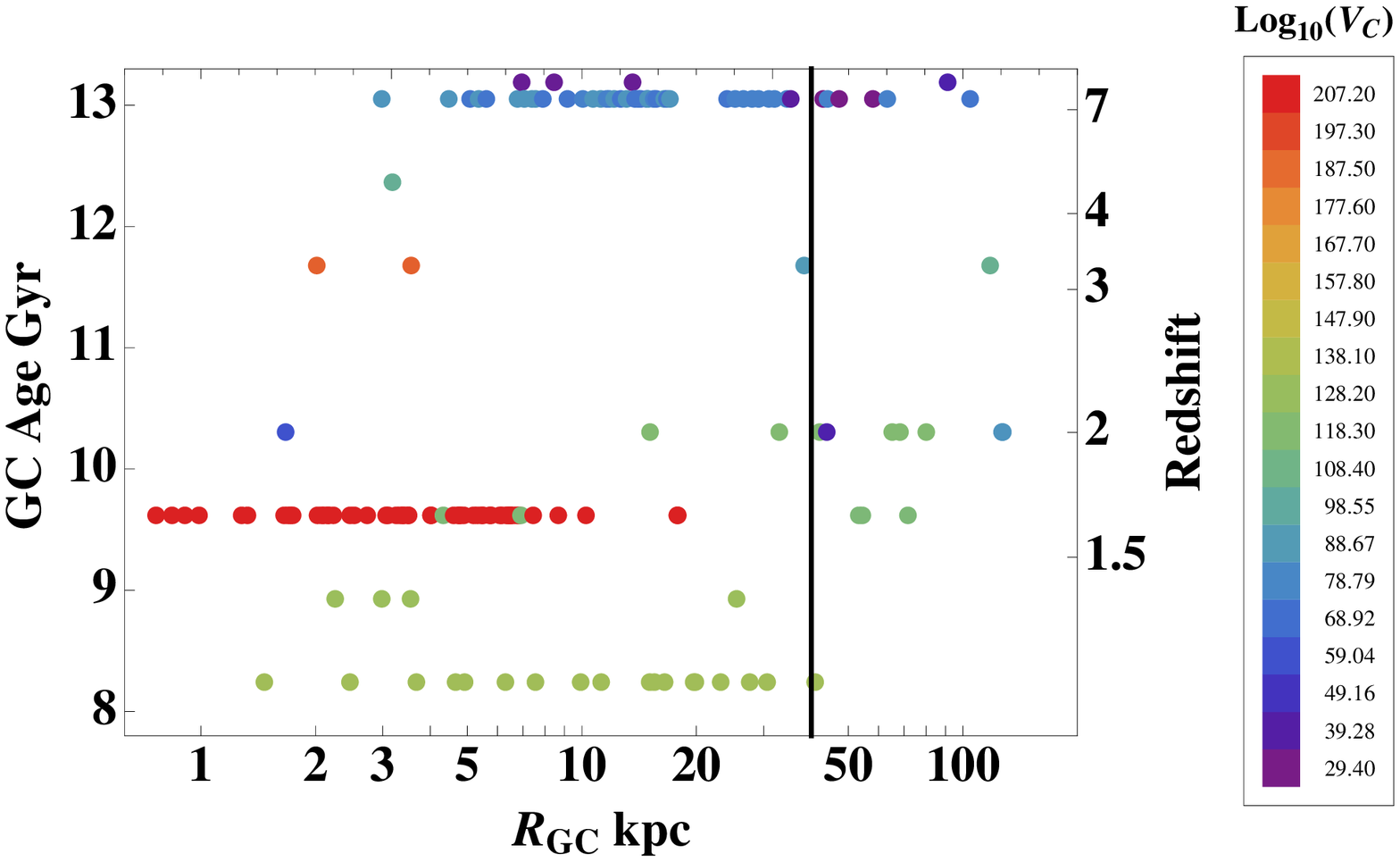,scale=0.42,trim= 180 0 500 0}\\
\epsfig{figure=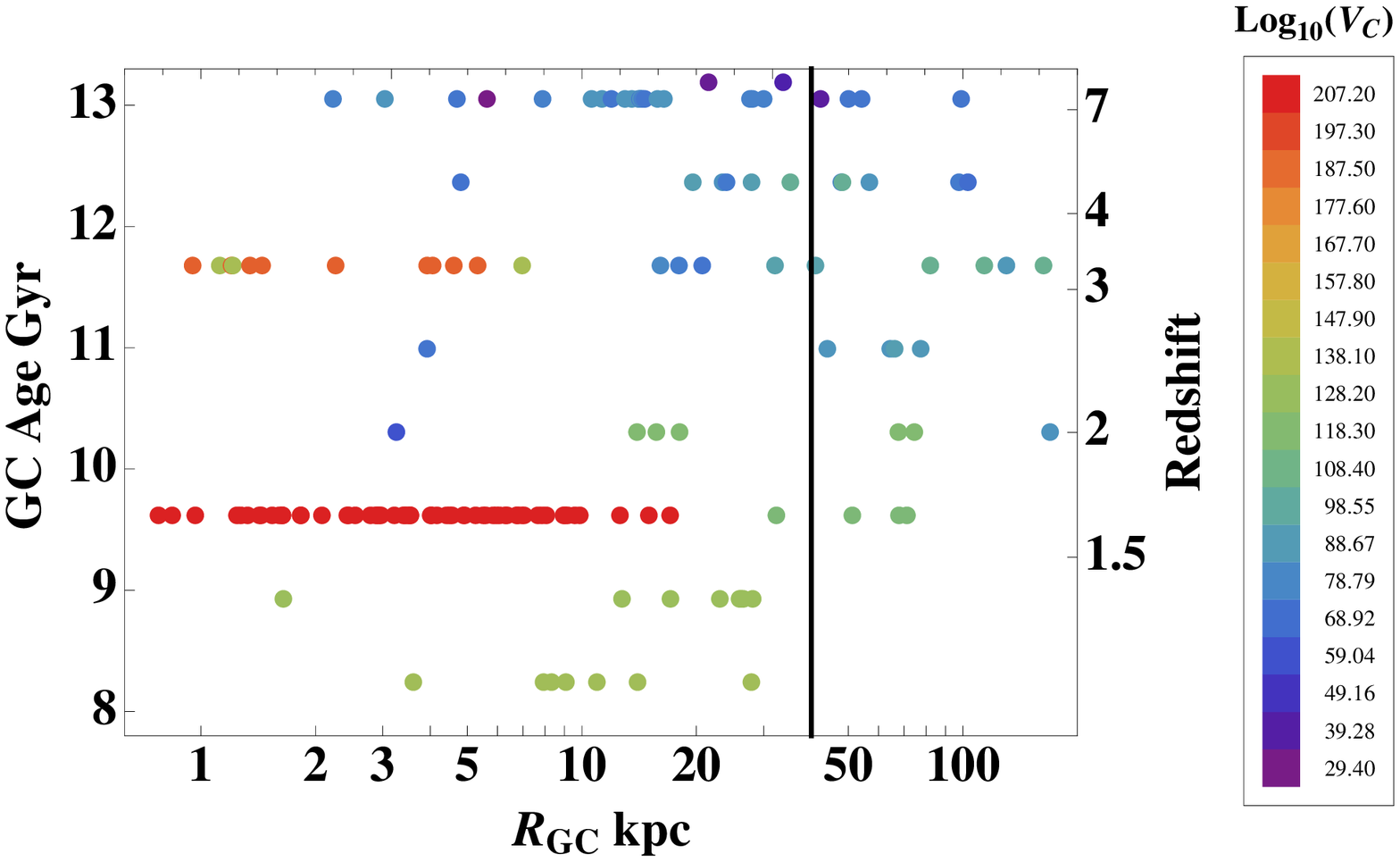,scale=0.42,trim= 280 0 50 0}\ \ \ \ \ \ \epsfig{figure=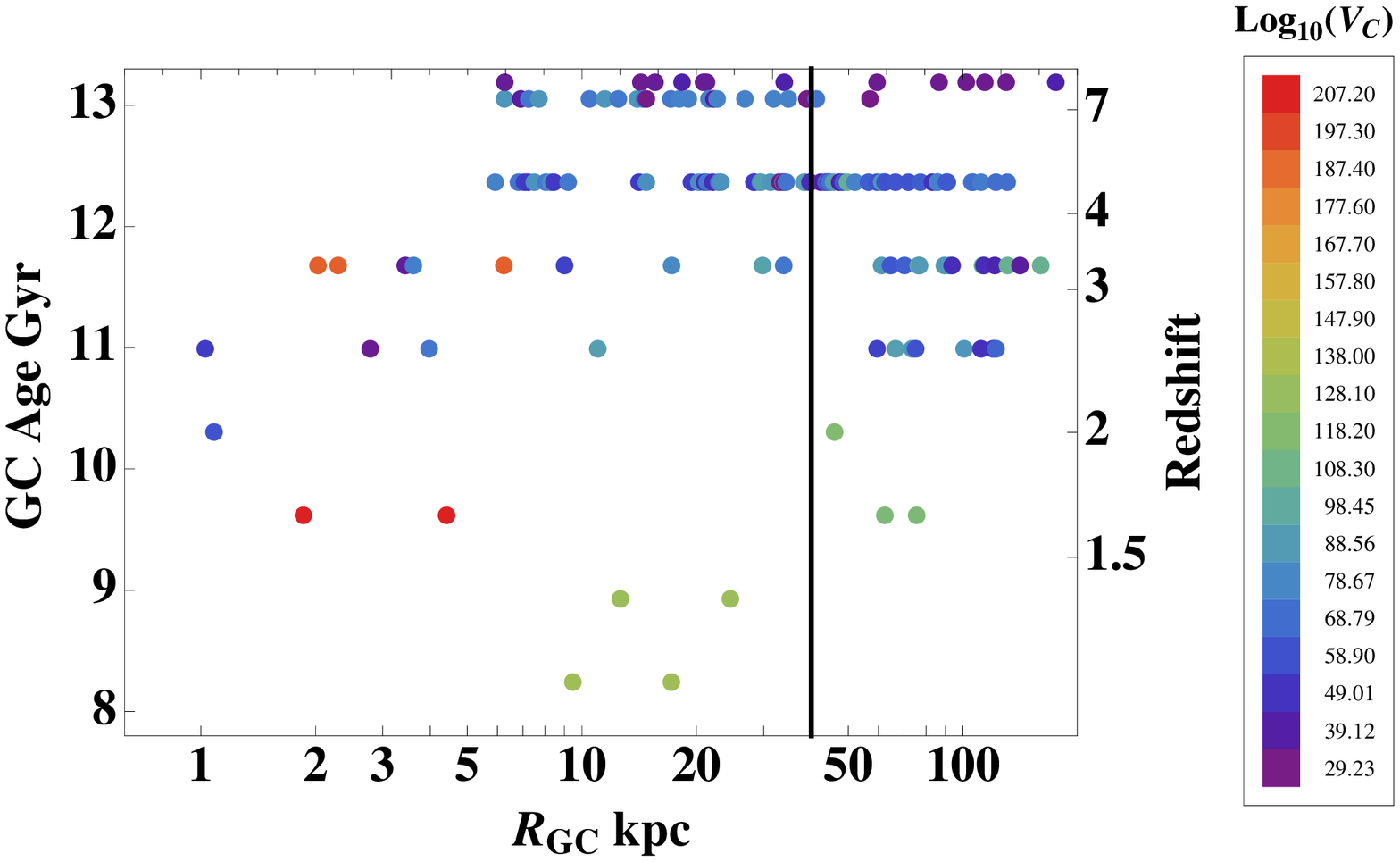,scale=0.42,trim= 180 0 500 0}
\caption{Galactocentric distances of the GCs versus the age in which
  they formed.  Each individual point represents one GC and the vertical line represents the radius which $\sim 95\%$ of the Milky Way GC population resides within. {\it Top
    Left.} CE Model. {\it Bottom Left.} KR13 Model. {\it
    Top Right.} KR13-bis Model. {\it Bottom Right.} Power Law Model}
\label{AgeR}
\end{figure*}

In order to improve the model, we explore which parameters in the
simulations have an effect on the CDF of Galactocentric distances of
the GCs.  $\Lambda$CDM predicts a hierarchical merging scenario for
the formation of galaxies, where the low mass haloes virialize first,
followed by the more massive galaxies at a later time.  In
Figure~\ref{AgeR} (top left), we plot the ages of the surviving
population of GCs (assumed to be the redshift of virialization of
their host halo) versus their Galactocentric distance at $z=0$.  The majority of the GCs that exist farthest from the center of the main halo formed at $2<z<6$.  On average, the halos which produce a larger fraction of GCs which are closer to the center versus farther away tend to be the haloes which either virialized before reionization at $z \simgt 6$
or after the redshift of virialization of the Milky Way at $z<2$.
This result can be understood as follows. GCs that formed in haloes at
$z\simgt 6$ belong to the most massive and rare haloes at those
redshifts (because there is a minimum halo mass threshold to form a GCs
system). Is well known that particles belonging to rare high-sigma
peak progenitors of the Milky Way today are preferentially found near
the center of the Milky Way, and so are their GCs systems
\citep{Diemand2005}. At the other extreme, dwarf haloes that virialize at very 
low redshift tend to be rather massive and thus their orbit decays
faster to the center due to dynamical friction (\eg, the Magellanic
clouds) and their GCs are preferentially deposited toward the halo
center.  In addition, GCs formed {\it in situ} at the redshift of
virialization of the Milky Way at $z \sim 2$ are centrally
concentrated, but this population is expected to have significantly
higher metallicity.
Indeed, the metal rich GCs tend to reside in the disk and closer to
the Galactic center while the metal poor GCs are preferentially found
in the halo \citep{Zinn1985}.  This supports the idea that at least a
fraction of GCs closer to the Galactic center formed in more massive
haloes since the metallicity of their host galaxy is proportional to
the mass of the halo.  Thus, we conclude that the CDF of the
Galactocentric distances of GCs in our simulation and the metallicity
distribution of the GC population are sensitive to the formation
efficiencies as a function of redshift and halo mass. Increasing the
formation efficiency of GCs at $z \sim 2-6$ will bias the CDF of the
radial distribution of GCs in our simulation towards larger radii,
whereas increasing the formation efficiency at $z<2$ and $z>6$ will
bias this function towards smaller radii. The effect of this
assumption will be explored in Section~\ref{subsec:aspen}.

The GCs that are very close to the galactic center are dominated by
{\it in situ} formation in the Milky Way progenitor. Reducing the
value of $\beta$ below the fiducial value $\beta=3.5$ will reduce the
fraction of {\it in situ} formation with respect to accreted GCs. The
effect of this assumption will be explored in
Section~\ref{subsec:aspen2}.

\begin{figure*}
\epsfig{figure=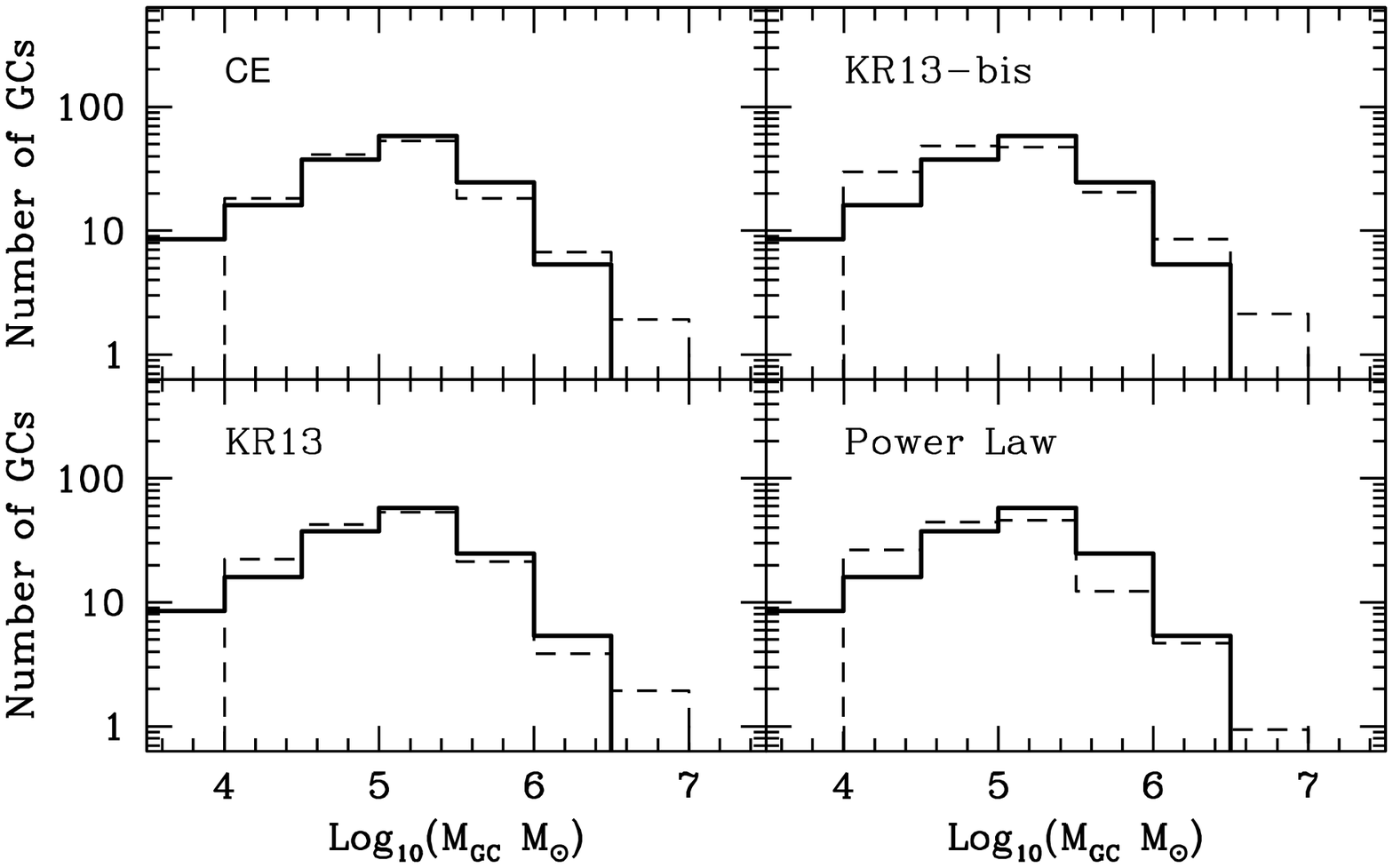,scale=0.8,trim= 200 0 200 200}
\caption{Mass function of GCs in each of our simulations (dashed
  lines) compared to that of the Milky Way (thick line).  The histograms have been normalized to a population of 150 GCs.  The peak of the mass function matches well for all models tested.}
\label{massfunc1}
\end{figure*}

In the top left panel of Figure~\ref{massfunc1}, we compare the GCMF
produced by the CE model to the Milky Way GCMF.  We see that the mass
function agrees quite well with the Milky Way GCMF and note that our GCIMF is, by construction, also guaranteed to
reproduce the GCMF of the local, isolated dwarf galaxies (see the right panel of Figure~\ref{minx2}).  This result is not trivial because the GCMF in the Milky Way is shaped by tidal effects in addition to two-body relaxation and stellar evolution.  The response of GCs to tides is dependent on the density of the GCs which was set to reproduce the effects of two-body relaxation in isolated local dwarf galaxies. 
Thus, we are confident that we are
capturing the relevant physical processes that determine the GCMF in
the Milky Way because any attempt to resolve a possible disagreement
adjusting the GCIMF or densities of GCs would break the
agreement with the GCMF in isolated dwarf haloes. This is an important conclusion
resulting from our self-consistent modeling of GC populations in
isolated dwarfs and the Milky Way.

A summary of the parameters of the CE model is in Table~1 and the
successes/failures of this model compared with observations are listed
in Table~2 with KS test probability statistics in Table~3.  

\subsection{The KR13 Model}
\label{subsec:aspen}

In order to improve the agreement of the model with the Galactocentric
distribution of GCs we relax the assumption of constant GC formation
efficiency that was assumed in the CE model. In this model we increase
the GC formation efficiency at high redshift, before reionization,
where the low mass dwarfs can survive until they reach the inner parts
of the progenitor of the main halo in the simulation.  Additionally,
we increase the efficiency at $1.5< z <2.5$ where the most massive
satellite haloes form their GCs.  The dashed black line in
Figure~\ref{formeff} shows $\eta_i(z)$ for this new model that we
refer to as the KR13 model, due to the two humps in the formation
efficiency over cosmic time. The formation efficiency in this model is
also in good agreement with the constraints on the GC formation
history derived from observations the galaxy luminosity functions in
the Hubble deep fields \citep{Katz2013}.

We produce a total of 150 GCs with this model.  In Figure~\ref{CDFs}
(left), we plot the CDF of the Galactocentric distances of GCs for the
KR13 model and note a much better agreement with the Milky Way than
what was found for the CE model (See Table 3).  The largest discrepancy between
observations and the model is at Galactocentric distances between
$20<R<60$~kpc.  

As stated before, the GCMF is largely robust against changes to the
formation efficiencies across cosmic time.  In the bottom left panel
of Figure~\ref{massfunc1}, we plot the GCMF for the KR13 model and
note that the peak is consistent with what we find for the Milky Way.
The radial velocities of the GCs in the KR13 model are closer to what
is observed in M31 as opposed to the Milky Way. Overall the kinematics
of GCs in our KR13 model are consistent with a Milky Way type spiral
galaxy.

Raising the formation efficiencies above their minimum values as we
have done in the KR13 model may overproduce GCs in isolated dwarf
galaxies.  We check to see if these formation efficiencies are
consistent with observations by comparing the specific frequencies
($S_N$) of the few isolated dwarf galaxies in our simulation to
observations.  Isolated dwarfs belong to the first two classes of GCs
which we described earlier, and although there are only a few isolated
dwarfs because the volume of the refined region in the simulation is
small, we see in Figure~\ref{SNa1}, that their specific frequencies
are consistent with what is expected from the observations of
\cite{Geo2010}\footnote{Masses of the haloes were converted into luminosities using equations 13 and 14 in \cite{Geo2010}.}.  A much larger volume is required to better understand
the dispersion in the individual values of $S_N$ for isolated dwarfs.

The host galaxy in which a GC forms impacts the chemical properties
seen in each individual GC.  The bimodal distribution of metallicities
of the Milky Way GC population is likely a reflection of accreted
dwarf galaxies which contributed GCs to the Milky Way as well as those
GCs which formed {\it in situ}.  Since the metallicity of a galaxy
scales with its luminosity, the GCs which formed in high mass galaxies
likely represent the high metallicity population while those which
formed in dwarfs contribute to the lower metallicity population.  The
KR13 model predicts that 41\% of the GC population formed {\it in
  situ} which may suggest that the Milky Way should exhibit a roughly
equal split of GCs with high and low metallicity.  In
Figure~\ref{metal}, we compare the expected metallicity distribution
of the accreted GCs formed in the KR13 model\footnote{Metallicities
  are calculated by using Equations~(17) and (18) from
  \cite{Muratov2010} and assuming the intrinsic metallicity
  distribution of GCs in each dwarf galaxy is lower than the metallicity of the host, in a half-normal distribution with a deviation of 0.4 dex. The results are
  largely independent of the shape of the intrinsic metallicity spread
  in the case of contribution from many dwarf galaxies. The same is
  not true for the GCs formed {\it in situ} in the Milky Way} 
 with the distribution of the Milky Way's GCs.  
Our simplistic model is used to demonstrate that the bimodal metallicity distribution of the Milky Way can be, in principle reproduced from a hierarchical merging scenario for the assumptions we have made on the intrinsic metallicity distribution.  We refer the reader to \cite{Tonini2013} where a much more in depth treatment of GC metallicities is presented within the context of the assembly of a large galaxy; however, the basic idea of an ``assembly scenario" is along the lines of the methodology used in this present work which has been shown to reproduce the bimodal properties of large galaxies.  \cite{Tonini2013} conclude that the distribution of the metallicities is dependent on the assembly and star formation history of the host galaxy.  It is unlikely that the assembly history of the main halo in the Via Lactea simulation exactly mimics that of the Milky Way.  The present calculation is used to demonstrate that a bimodal population can be reproduced in our present framework and that it is likely also sensitive the the masses of the haloes which contribute GCs since the metallicity of the stars in a halo is partially dependent of the mass of the halo.

It is clear that this model produces significantly fewer low metallicity GCs than
expected and it is unlikely that those GCs formed {\it in situ} can
account for the deficit at low metallicities.  While we have
successfully reproduced the radial distribution of GCs as well as
multiple other characteristics of the Milky Way GC population, this
model fails to reproduce the metallicity distribution seen in the
Milky Way which has a surviving population likely dominated by
very old accreted GCs.

\begin{figure*}
\epsfig{figure=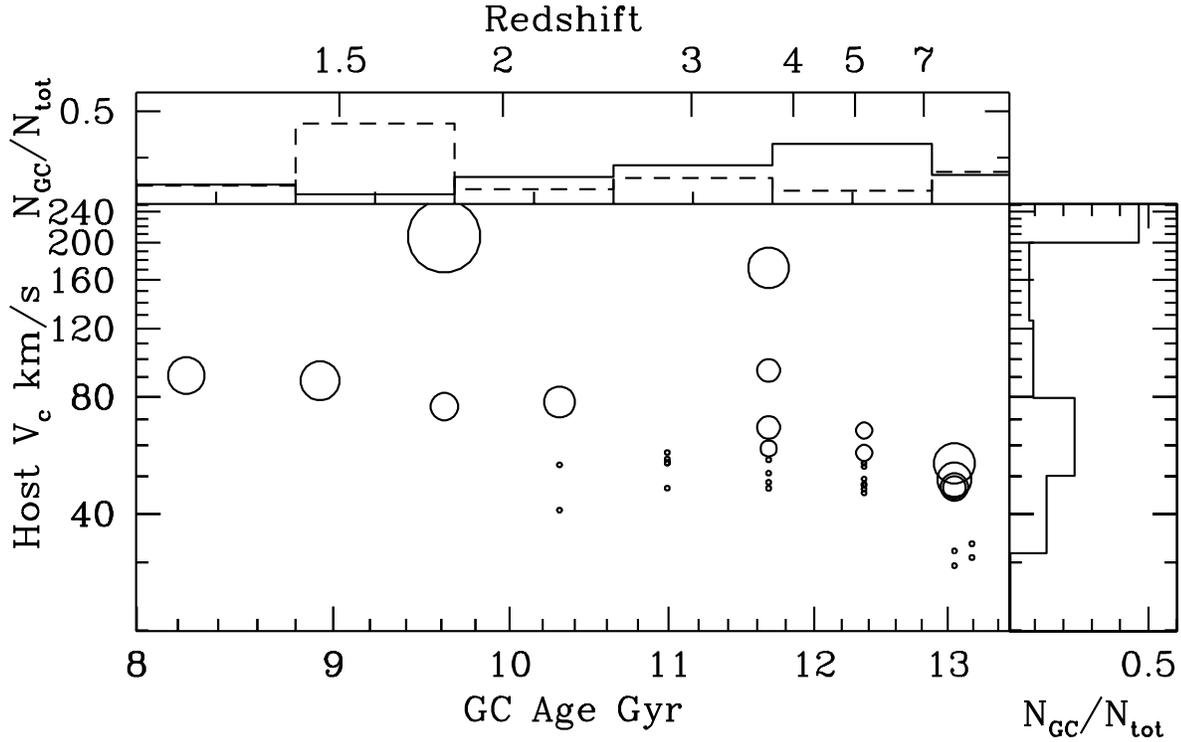,scale=0.8,trim= 200 50 200 180}
\caption{Formation epoch of GCs versus the circular velocity of the
  host halo for the KR13 model.  The radius of the circle is
  proportional to the logarithm of the number of surviving GCs each
  halo contributed to the final population of the main halo.  The top
  panel is a histogram of the ages of the GCs in the simulation
  (dashed line) compared to the ages of 93 of the Milky Way GCs (solid
  line) compiled by \protect\cite{Forbes2010}.  The right panel is a
  histogram of the number of haloes of a given $V_C$ which contributed
  surviving GCs.  The large bubble at the top left represents the main
  halo in the simulation.  }
\label{abub}
\end{figure*}

\begin{figure}
\epsfig{figure=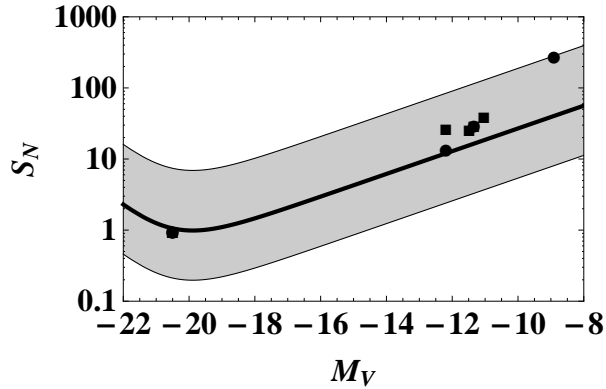,scale=1.1}
\caption{Specific frequency of galaxies in our simulation versus their
  absolute magnitude.  The circles represent isolated dwarf haloes in
  the KR13 model, and the squares represent the isolated haloes in the
  KR13-bis model. The points at $M_V=-20.5$ represent the
  specific frequency for the main halo in our simulation which we
  assume to have an absolute visual magnitude equal to that of the
  Milky Way.  The shaded region represents the expectation from
  \protect\cite{Geo2010} with $10^{-5}<\eta<3.5\times10^{-4}$.  The
  thick black line in the middle of the shaded region represents
  $\eta=5.5\times10^{-5}$ as derived by \protect\cite{Geo2010}.}
\label{SNa1}
\end{figure}

Since we have a two peaked model for the formation efficiencies, one
might expect that the age distribution of the GCs in our simulation
also shows this bimodal characteristic.  In the top panel of
Figure~\ref{abub} we plot a histogram of the ages of the GCs in our
simulation (solid line) and compare to those known for the Milky Way
GC population (dashed line).  The ages of most Milky Way GCs are only
known to a precision of $\pm 1$~Gyr [but see \cite{Katz2013}] and any
underlying bimodality in the age distribution is smoothed out by these
large uncertainties.  Furthermore, our model assumes that all GCs in
an individual galaxy form synchronized in an instantaneous burst,
neglecting any intrinsic age spread. This simplifying assumption is
reasonably realistic for dwarf galaxies (because of their short
dynamical time scale) but is likely less realistic for GCs formed {\it
  in situ} in the Milky Way.

In order to test whether the ages of GCs in our simulations agree with
those of the Milky Way GC population, we convolve the ages of GCs in
our simulation with a Gaussian distribution with a standard deviation
of 1~Gyr. Figure~\ref{Ages} compares the results with the observed
ages of GCs in the Milky Way.  We can see that simulated GCs from the
KR13 model is not double peaked and appears as a continuous formation
scenario consistent with what is seen for the 93 Milky Way GCs with
age estimates \citep{Forbes2010}.  There is still a overabundance of
GCs forming at the time of virialization of the Milky Way, but as
mentioned before, one should convolve the GCs formed {\it in-situ} in
the Milky Way with a larger spread for their age distribution.

\begin{figure*}
\epsfig{figure=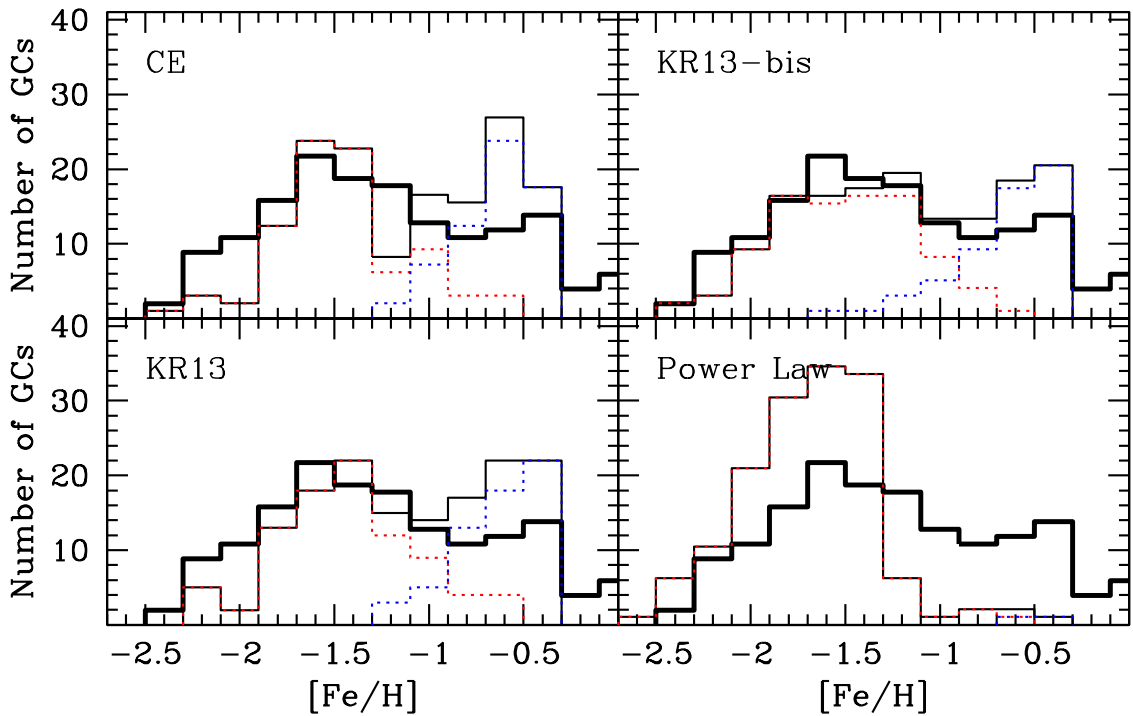,scale=.8,trim= 200 0 200 200}
\caption{Metallicities of GCs in our simulations (thin line) are
  compared to what is observed for the Milky Way (thick line).  The GCs which formed {\it in situ} are shown in blue and the GCs which were accreted are shown in red.
  $\sim30\%\ (\sim60\%)$ of Milky Way GCs have [Fe/H] $>-1\ (-1.5)$.}
\end{figure*}

\begin{figure*}
\epsfig{figure=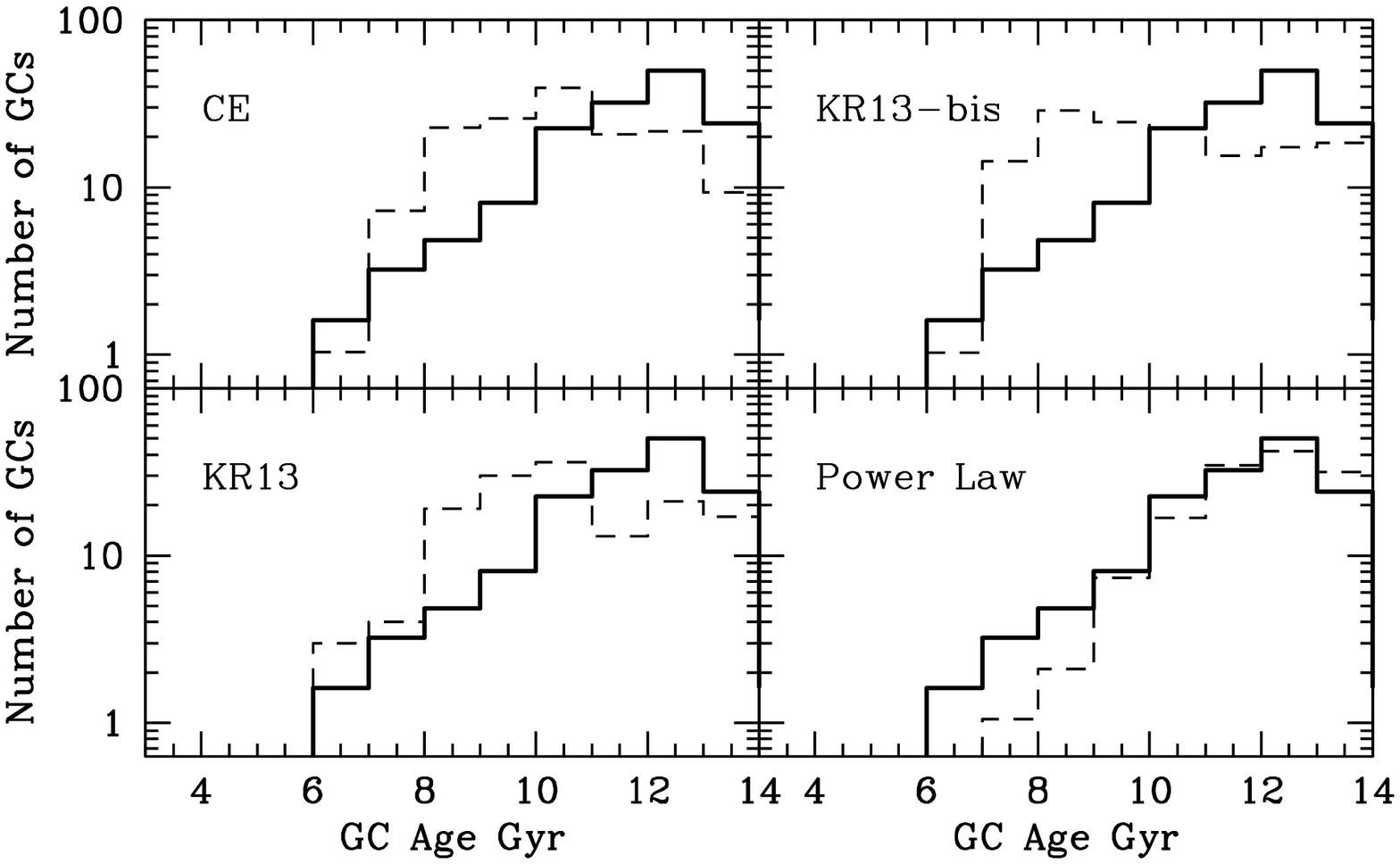,scale=.8,trim= 200 0 200 200}
\caption{Ages of GCs in each of our simulations randomized over a
  Gaussian distribution with $\sigma=1$ Gyr (dashed lines) compared to
  ages of 93 of the Milky Way GCs (solid line) compiled by
  \protect\cite{Forbes2010}.  The histograms have been normalized to a
  population of 150 GCs.}
\label{Ages}
\end{figure*}
A summary of the parameters of the KR13 model is in Table~1
and the model's successes/failures in matching observations can be
found in Table~2 with KS test probability statistics in Table~3.

\subsection{The KR13-bis Model}
\label{subsec:aspen2}

As previously discussed, the metallicity of GCs is likely determined
by the mass of the host galaxy in which they form.  A shortcoming of
the KR13 model is the deficiency of low metallicity GCs with respect
to observations. In order to produce more low metallicity GCs, we can
increase the number of GCs formed in small mass haloes at high redshift with respect to
those formed in larger mass haloes.  We test a third model, the KR13-bis
model where we adopt $\beta=3$ (\ie, $\gamma=-0.14$) in
Equation~(\ref{oldeqn}) so that the model produces more GCs that
formed in dwarf galaxies and were later accreted onto the Milky Way,
rather than formed {\it in situ} in the Milky Way.

We show $\eta_i(z)$ for the KR13-bis model as the dotted line in
Figure~\ref{formeff}.  We can see in the top right panel of
Figure~\ref{metal} that the metallicity distribution for this model is
significantly improved over the KR13 model and in this model, only
$38\%$ of the total surviving GC population formed {\it in situ} in
the Milky Way.  Although the fraction that formed {\it in situ} is only slightly lower, we form more GCs in older lower mass halos which improves the metallicity distribution.  The accreted GC population dominates the low
metallicity peak but also contributes some higher metallicity GCs as
also found by \cite{Muratov2010}.  In this model number of GCs formed
{\it in situ} in the Milky Way is roughly what is needed to fill the
gap at the high metallicity end of the distribution.  The GCMF
produced by the KR13-bis model (see top right panel of
Figure~\ref{massfunc1}) remains largely unchanged from the the KR13
models and the CDF of the radial velocities is identical to that of
the KR13 model and to observations of M31 (see Figure~\ref{CDFs}
(right)).  Likewise, the CDF of the Galactocentric distances of the
GCs is quite consistent with what is seen in the Milky Way, similarly
to the KR13 model (see Figure \ref{CDFs} left).

\begin{figure*}
\epsfig{figure=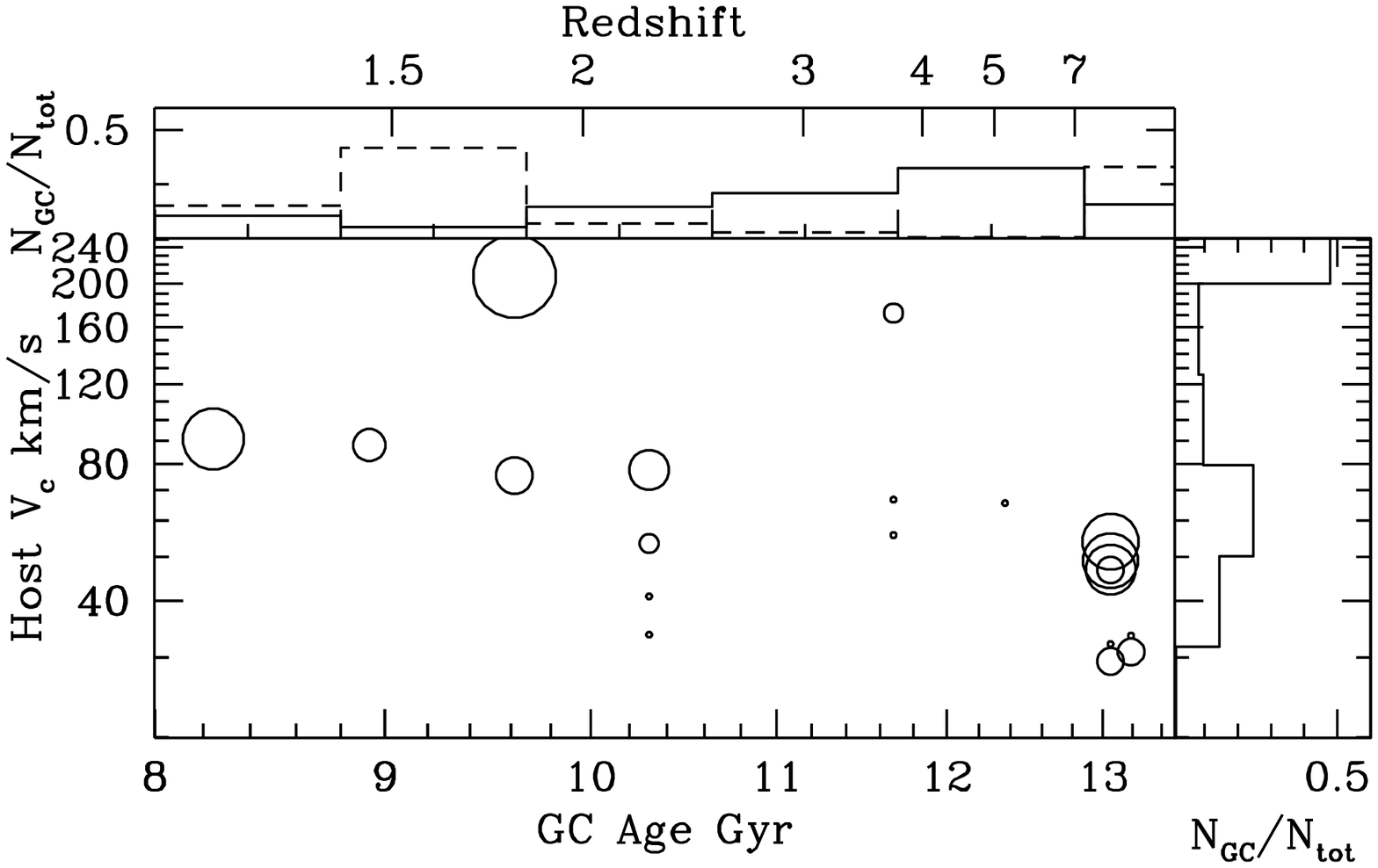,scale=0.8,trim= 200 50 200 180}
\caption{Formation epoch of GCs versus the circular velocity of the
  host halo for the KR13-bis model.  The radius of the circle is
  proportional to the logarithm of the number of surviving GCs each
  halo contributed to the final population of the main halo.  The top
  panel is a histogram of the ages of the GCs in the simulation
  (dashed line) compared to the ages of 93 of the Milky Way GCs (solid
  line) compiled by \protect\cite{Forbes2010}.  The right panel is a
  histogram of the number of haloes of a given $V_C$ which contributed
  surviving GCs.  The large bubble at the top left represents the main
  halo in the simulation.}
\label{a2bub}
\end{figure*}

Since decreasing $\beta$ in Equation~(\ref{oldeqn}) effectively
increases the formation efficiency of GCs in lower mass galaxies, a
larger proportion of GCs that survive to $z=0$ have formed earlier in
low mass galaxies (see Figure \ref{a2bub}).  The most massive dwarf
galaxies at high redshift contribute significantly to the total
accreted GC population, accounting for about $40$\% of surving GCs in
the Milky Way.  Adopting a model with such high efficiencies may
overproduce the number of GCs in isolated dwarf galaxies.  In
Figure~\ref{SNa1} we compare the specific frequency of GCs in isolated
dwarfs in the KR13-bis model (squares) to observations.  While there
are only a few isolated haloes which have GCs that have survived until
the present, these few galaxies fall perfectly within the range
observed in local dwarf galaxies.

A summary of the parameters of the KR13-bis model is in Table~1
and the model's successes/failures in matching observations can be
found in Table~2 with KS test probability statistics in Table~3.

\subsection{The Power Law Model}
\label{subsec:powerlaw}

We test one final model where GC formation is extremely biased to
occur in high redshift dwarf galaxies.  We adopt a value of
$\beta=1.5$ (\ie, $\gamma=-0.57$) in Equation~(\ref{oldeqn}) and
adjust the formation efficiency as a function of redshift to a power
law that is $\propto (1+z)^2$ ($\eta_i(z)$ for the Power Law model is
the long dashed line in Figure~\ref{formeff}).  This model seeks to
determine if the Milky Way's GC population could be formed entirely
from an accreted population.  Of the 143 GCs that survive within the
main halo until $z=0$, only two of these GCs were formed {\it in situ}.

As was observed in the three previous models, the mass function and
the CDF of the GC's velocities remains largely unchanged and are
consistent with what we expect for a Milky Way type galaxy.
Furthermore, the ages of the GCs formed in this simulation agree well
with the measured ages for the Milky Way GCs.  However, we find that
the CDF of the positions of the GCs drastically under predicts what is
seen in the Milky Way and the GCs in this simulation are biased to
much farther distances.  Since we adopted such an extreme value of
$\beta=1.5$ and thus the majority of the surviving GC population was
accreted from high redshift dwarf galaxies, we also find that we
significantly over predict the number of low metallicity GCs in the
Milky Way.

A summary of the parameters of the Power Law model is in Table~1 and
the model's successes/failures in matching observations can be found
in Table~2 with KS test probability statistics in Table~3.

\begin{table*}
\label{one}
\centering
\begin{tabular}{@{}lccccccccccc@{}}
Model & $N_{GC}^{tot}$ & $N_{GC}^{acc}$ & $N_{GC}^{surv}$ & $N_{GC}^{in-situ}$ & $N_{Dw}^{acc}$ & $N_{Dw}^{surv}$ & $f_N^{surv}$ & $f_M^{surv}$ & $N_{GC}(z>7)$ & $f_M^{surv}(z>7)$\\
\hline
CE & 145 & 335 & 84 (58\%)& 61 (42\%) & 63 & 43 & 27\% & 20\% & 5(3\%) & 10\%\\ 
KR13          & 150 & 279 & 89 (59\%)& 61 (41\%) & 52 & 38 & 30\% & 19\% & 26(17\%) & 15\%\\
KR13-bis      & 146 & 238 & 90 (62\%)& 56 (38\%) & 32 & 21 & 36\% & 20\% & 48(33\%) & 22\%\\ 
Power Law     & 143 & 301 & 141(99\%)& 2  (1\%)  & 100& 70 & 46\% & 31\% & 36(25\%) & 24\%\\
\hline
\end{tabular}
\caption{$N_{GC}^{tot}$ is the total number of GCs that survived to
  $z=0$. $N_{GC}^{acc}$ is the total number of accreted GCs.  $N_{GC}^{surv}$
  is the number of accreted GCs that survived to $z=0$.  $N_{GC}^{in-situ}$
  is the number of surviving GCs that formed {\it in situ}.
  $N_{Dw}^{acc}$ is the number of haloes that contributed GCs to the Milky
  Way halo, $N_{Dw}^{surv}$ S is the number of haloes that contributes
  surviving GCs to the Milky Way halo.  $f_N^{surv}$ is the percentage
  of GCs that survives by number and $f_M^{surv}$ is the percentage of
  GCs that survive by mass. $N_{GC}(z>7)$ and $f_M^{surv}(z>7)$ are the total number of GCs that form at redshifts $z>7$ and the fraction of the mass that survives from those redshifts. }
\end{table*}

\begin{table*}
\label{two}
\centering
\begin{tabular}{@{}lccccccc@{}}
Model & GC Positions & GC Velocities & Peak Mass & Metallicity & Ages & $S_N$\\
\hline
CE & X & $\surd$ & $\surd$ & - & $\surd$ & $\surd$\\ 
KR13 & $\surd$ & $\surd$ & $\surd$ & X & $\surd$ & $\surd$\\
KR13-bis & $\surd$ & $\surd$ & $\surd$ & $\surd$ & $\surd$ & $\surd$\\ 
Power Law & X & $\surd$ & $\surd$ & X & $\surd$ & $\surd$\\ 
\hline
\end{tabular}
\caption{Success of each of the three simulated models compared with
  observations.  ``$\surd$" represents a potential agreement, ``X"
  represents a clear disagreement, and ``-" means the potential
  agreement is unclear.}
\end{table*}

\begin{table*}
\label{three}
\centering
\begin{tabular}{@{}lccc@{}}
Model & GC Positions & GC Velocities (MW) & GC Velocities (M31)\\
\hline
CE & 0.11\% & 0.41\% & 30\%\\
KR13 & 4.8\% & 0.47\% & 46\%\\
KR13-bis & 2.9\% & 0.32\% & 60\%\\
Power Law & 0\% & 0.97\% & 75\%\\ 
\hline
\end{tabular}
\caption{KS Test probabilities for each of the models compared with observational results.}
\end{table*}

\section{Discussion}

In order to match the observed metallicity distribution function,
Galactocentric distances, and ages of GCs in our models, we find that
the specific (per unit dark matter halo mass) formation efficiency of
GCs, $\tilde \eta_i$, should be bimodal with a peak at redshift $z>6$
and should be higher in low mass haloes than in larger mass haloes. 

It remains unclear whether the second peak of the formation efficiency
at low redshift is a robust result of the model. For instance, a low
redshift increase of the formation efficiency may be produced by GCs
that may form as a result of ``minor'' mergers with the Milky Way
after its virialization, a GC formation process that we have not
modeled.  However, the GC population we have designated as forming
{\it in situ} within the Milky Way may be interpreted as a
population that formed as a result of major galaxy mergers which coincide
with rapid mass growth and therefore virialization of the halo.  In
order to reproduce the observable characteristics of the Milky Way GC
population, some significant portion of GCs (about 30\%) should form
with the Milky Way, whether it be {\it in situ} or in mergers.

We can see from the left panel of Figure \ref{CDFs} that the positions of GCs in the Milky Way exhibit a strong radial gradient within $\sim10$ kpc and our models fails to reproduce this property.  From the top left panel of Figure \ref{AgeR} it is also clear that the majority of GCs which populate this region were formed {\it in situ}, within the main halo of the simulation.  Unlike the accreted GCs where the initial conditions for their orbits are taken directly from the kinematics and positions of the accreted dwarf haloes, the initial conditions for the {\it in situ} population of GCs are relatively unconstrained.  The constant density sphere which we have assumed did not reproduce the radial gradient exhibited by Milky Way GCs and therefore, the initial conditions we have assumed are unlikely to represent to true initial conditions of the {\it in situ} population.  In all models (excluding the Power Law model), the population of GCs within 10 kpc is also dominated by the GCs that formed {\it in situ} and as we can see in left panel of Figure \ref{CDFs}, these models are identical up to $\sim10$ kpc (once again excluding the Power Law model).  

The KS test probabilities for the positions will significantly improve if we choose initial conditions for the {\it in situ} population that reproduce this radial gradient.  We would also likely have to increase the formation efficiency of the {\it in situ} population as more GCs will be placed closer to the center of the main halo which leads to a high probability that they will encounter the bulge.  The CE models begins to diverge from the double peaked models after this fiducial radius which represents the point at which the accreted population of GCs represents the majority of the population.  This suggests that the major differences between the positions in these models is due to the populations of accreted GCs and not the GCs that formed {\it in situ}.  Since the initial conditions of the {\it in situ} population are not known, simply inserting a radial gradient initially, although it will improve the KS test probabilities for the positions, does not provide any additional information, as this would require testing multiple different initial conditions in order to determine which are compatible with the observed radial gradient.  Given the results of our simulations, it is unlikely that the {\it in situ} population formed in a constant density sphere because we have failed to reproduce the radial gradient.  For these reasons, we stress that the KS test probabilities for the positions should be compared between models in order to determine which model provides a better fit to the observational data.

Since we have only tested one halo, with one choice of disk and bulge parameters, it is important to understand how deviations from these parameters might effect our simulations.  A more massive bulge will certainly lead to more tidal destruction as the sphere of influence will become larger.  However, as we described previously, we are tightly constrained by observations on how massive our classical bulge can be.  Our simulations are likely much less sensitive to changes in parameters for the disk as long as it does not become so dense that it can tidally disrupt the GCs in our simulations.  This is unlikely to be the case because there exists a population of GCs in the Milky Way which live close do the disk and if the disk could significantly disrupt the GCs, this population would not exist.  Furthermore, disk shocking only minimally effects the GC population in our simulation and therefore, small deviations about the chosen parameters are unlikely to result is disk shocking becoming a dominant effect.  The main role that changes to the disk parameters may cause is changes to the velocities of GCs.  For GCs outside the disk, $V_c\propto \sqrt{M}$, and since our $z=0$ disk mass is $\sim70\%$ greater than the disk mass estimated by \cite{Bovy2013}, we can expect the velocities of our GCs might be $\sim30\%$ larger than what is observed.  We can see in the right panel of Figure \ref{CDFs} that the GCs in our simulation tend to have higher velocities than Milky Way GCs and that they tend to agree with M31 which is likely to have a more massive disk than the Milky Way.  Decreasing our disk mass will almost certainly relieve some of the tension in the CDF of the velocities.

Our model makes distinct predictions for the number of GCs in the
surviving population which were accreted versus formed {\it in situ},
within the Milky Way.  We found that in order to match the observed
metallicity distribution of Milky Way GCs, the majority of the
surviving population (about $62$\%) has to form in lower mass high redshift
satellite haloes.  \cite{Forbes2010} studied 93 of the Milky Way's
$\sim 150$ GCs and provide a lower limit for the number of Milky Way
GCs that were accreted and found that $27-47$ GCs ($\approx 30\%-50\%$
of the population) were accreted from $6-8$ dwarf galaxies.  Our most
successful model, the KR13-bis model, predicts that $62\%$ of
the Milky Way's GC population was accreted which is slightly higher
than the upper limit from \cite{Forbes2010}.  Our KR13-bis simulation
predicts that the present population of accreted GCs (90 GCs) comes
from a total of $19$ dwarfs, with $9$ dwarfs each contributing less
than 3 GCs (for a total of 11 GCs) and $10$ contributed at least 3
surviving GCs each (for a total of 79 GCs). The number accreted dwarfs
is slightly higher than the lower bound suggested by
\cite{Forbes2010}.  However, looking at the metallicity distribution
of GCs in the Milky Way, $\sim 30\%\ (\sim 60\%)$ of Milky Way GCs have
[Fe/H] $>-1\ (-1.5)$ and the KR13-bis model predicts a reasonable number of accreted GCs, within this range, that produce the metal poor
population.

Because we use only one realization of the Milky Way halo (the VL~II
merger tree), we are not able to capture the variance of the GC
distribution due to different merger histories. We point out that even
with the given merger tree, the stochasticity due to the assignment of
GCs masses extracted from the GCIMF produces some fluctuations in the
model results. Thus, we do not expect to reproduce exactly the Milky
Way's GC population. Rather, our goal has been to understand the dominant
physical processes that determine the various observables.

Despite our inability to determine the process which form GCs within
the main halo, the observed ages of GCs in the Milky Way constrain the
approximate epochs of formation of GCs. This allows us to make
predictions for the destruction rates of GCs within the main halo.
The KR13-bis model predicts a destruction rate by mass of
$\sim 80\%$.  We point out that this value fluctuates considerably
between different runs of the same simulation but never goes below $\sim65\%$.  The destruction percentage in our simulations is
very sensitive to what happens to the few $10^7$~M$_\odot$ GCs in our
simulation. Because we aim to form $\sim 150$ GCs in our model, we
expect only a few of the highest mass GCs to form in the entire
simulation and therefore, a large fluctuation in destruction
percentage is not unexpected.

\begin{figure*}
\epsfig{figure=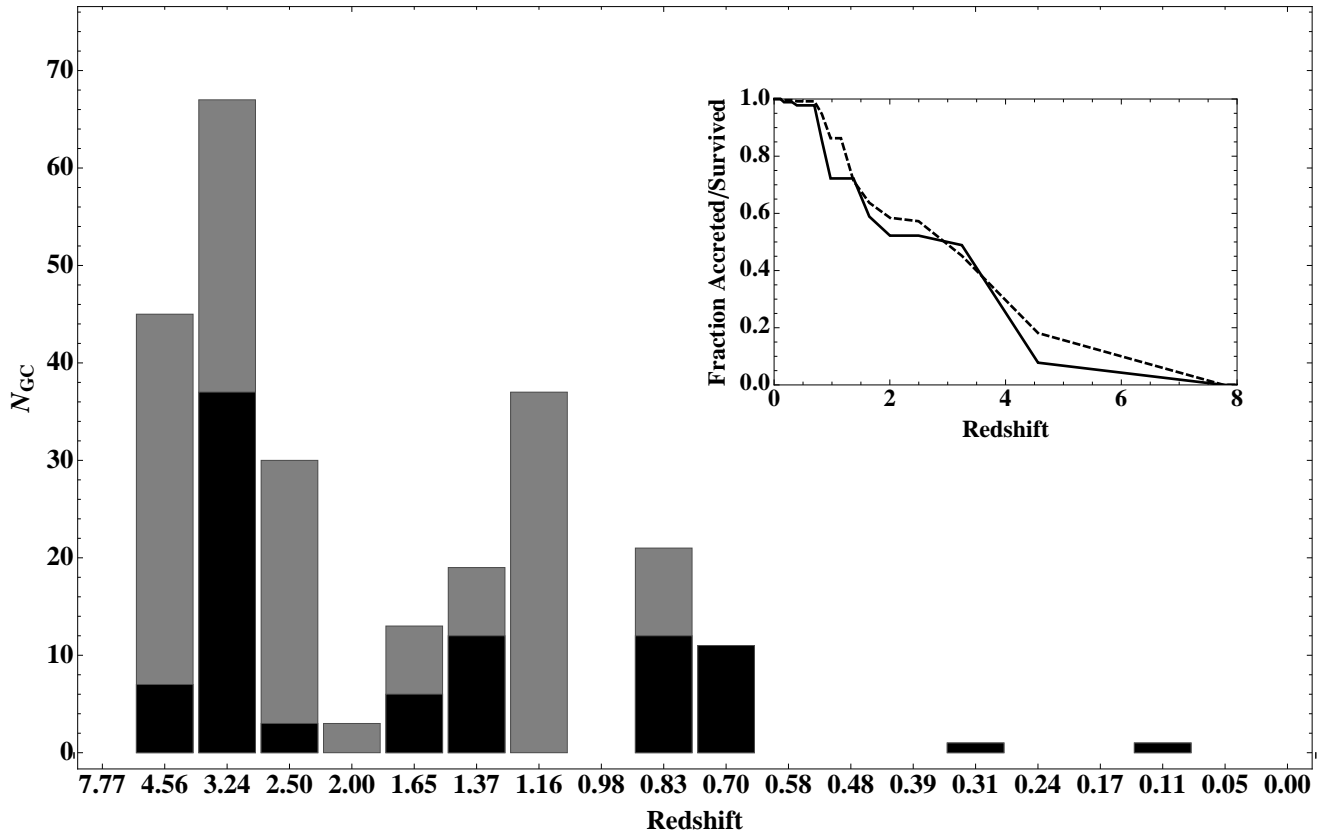,scale=.7}
\caption{The grey bars represent the number of GCs that were accreted
  as a function of redshift while the black bars represent the number
  of GCs that survive.  The solid line in the inset shows the CDF of
  the number of GCs that survived as a function of the redshift they
  were accreted while the dashed line represents the CDF of the number
  of GCs accreted as a function of redshift.}
\label{funhist}
\end{figure*}

In Figure \ref{funhist} we plot the number of GCs that are accreted as
a function of redshift for out best fit KR13-bis model as well as the
number that survived.  Although $33\%$ of GCs form in dwarf haloes at $z>7$, we can see from the inset of Figure \ref{funhist} that the bulk of all surviving GCs were accreted onto the main halo at $z<4$ with about half having been accreted between $z=2$ and $z=0.7$. 

We have also neglected the idea of a generic ``infant mortality" of GCs where
stars clusters, independent of mass, are destroyed within a few tens
of Myr \citep{Bastian2005}.  This destruction mechanism has no effect
on our simulations ability to reproduce the observable characteristics
of the Milky Way's GC population if it is independent of mass;
however, this certainly may influence the role of GCs from a
cosmological standpoint.  Another dynamical process that we have
neglected is the effect gas expulsion, suggested by
\cite{Baumgardt2008}, which may destroy up to $95\%$ of all clusters
within a few tens of Myr of formation.  This effect is dependent on
the initial mass of the cluster and tends to destroy the lower mass
end of the GCIMF more effectively, thus shaping the surviving mass
function.  \cite{Baumgardt2008} have claimed that this destruction
mechanism is nearly independent of the external tidal field and should
therefore affect GCs in dwarf galaxies as well as in larger Milky Way
type galaxies.

Since GCs emit the majority of their
ionizing radiation within $\sim 10$~Myr of their formation, these infant
mortality mechanisms, which destroys GCs on slightly longer time scales, do not
prohibit the evolution of the massive stars within GCs.
\cite{Katz2013} found that assuming a destruction percentage by mass
of $\sim 90\%$ for GCs, if $\sim40\%$ of GCs formed before $z\sim
6$, then they would have played a major role in the reionization of the
Universe.

The destruction percentage predicted in our simulations is also
sensitive to the upper and lower limits of the masses in the GCIMF. We
have assumed $M_{low}=10^5$~M$_\odot$, but if we lower this value, the
destruction rate will increase roughly as given in
Equation~(\ref{eq:fM}) (\ie\ by a factor $\sim 1.4$ for
$M_{low}=10^4$~M$_\odot$), which increases the lower limit of the destruction rate by mass in our simulations 
from $\sim65\%$ to $\sim91\%$.  

The destruction percentage in our simulations is further sensitive to our choice of $x_{crit}$ which is the ratio between the half light radius and the tidal radius for which we consider our GCs destroyed.  Given the density of our GCs, which was derived to match the mass function of GCs in dwarf galaxies, this destruction mechanism is only relevant in the very inner parts of the galaxy; however, since many of the GCs fall in on radial orbits, many GCs have at least one passage close to the center over their many Gyr lifetimes.  Observations of the Milky Way population do not show any GCs residing within a few hundred parsecs of the galactic center where this process is effective.  

We tested the effects of turning this process off for the KR13-bis model and found that, despite producing many more GCs and found a destruction percentage by mass of $\sim50\%$ compared to the previous $80\%$.  However, the velocity distribution, mass function, and ratio between the number of GCs formed {\it in situ} versus accreted which survive also remained the same which gives the right metallicity distribution and ages.  The positions of GCs in the simulation compared to the Milky Way slightly improve because more GCs are allowed to remain at the center.  This suggests that if we renormalize the efficiency as a function of redshift to produce the correct number of GCs, all aspects of the KR13-bis model which reproduce observations of the Milky Way GC population will once again be reproduced with a smaller destruction percentage.  Thus we emphasize that the choice of normalization for the efficiency of forming GCs as a function of redshift is degenerate with this destruction mechanism as well as our choice of lower limit on mass in addition to any assumption we make for infant mortality.  We are constrained to not overproduce or underproduce GCs in isolated dwarf galaxies and since our choice of $x_{crit}$ is slightly low, the fact that we reproduce the populations of GCs in isolated dwarfs suggests that our choice of normalization, although degenerate with many parameters, is reasonable since our choice of $x_{crit}$ does not effect the populations of GCs in the isolated dwarfs.  To conclude this point, the shape of the efficiency curve as a function of redshift produces the correct positions, mass function, velocity distribution, metallicity distribution and age distribution and this property is entirely robust to changes in the normalization of the curve.

\section{Did Globular Clusters Reionize the Universe?}
Using the constraints from our simulations, we can determine the role that GCs may have played in the reionization of the Universe.  \cite{Ricotti2002} demonstrated that the expected number of ionizing photons per baryon by GCs per Hubble time at redshift 7 is:
\begin{equation}
N_{ph}^{gc}=f_{esc}\eta\omega_{gc}\frac{t_H(z=7)}{\Delta t_{gc}}.
\end{equation}
where $f_{esc}\sim1$ is the escape fraction of photons from GCs (see \cite{Ricotti2002} for discussion of this value), $\eta=8967$ is the number of ionizing photons per baryon in stars which is constrained from STARBURST99 \citep{sb99} which can certainly increase with a more top-heavy stellar IMF as discussed in \cite{Schaerer2011}, $t_H(z=7)$ is the Hubble time at $z=7$, $\omega_{gc}\approx2.1f_{di}(2.7^{+2.3}_{-1.7}\times10^{-4})$ is the fraction of cosmic baryons converted into stars, $\Delta t_{gc}$ is the time period over which GCs form, and $f_{di}$ is the inverse of the survival percentage by mass.  From Table 1, we know that in the KR13-bis model predicts $f_{di}\approx4.55$ during the reionization epoch.  Furthermore, the value of $t_H(z=7)/\Delta t_{gc}$ effectively sets the fraction of GCs which form prior to the epoch of reionization, $f_{ri}$, which is found in our KR13-bis model to be $f_{ri}=29\%$.  We can, therefore, rewrite the previous equation as follows:
\begin{equation}
N_{ph}^{gc}=5.08f_{ri}f_{di}=6.7
\end{equation} 

As previously discussed, our simulations likely predict the correct shape of $\eta_i$  as a function of redshift, but the normalization is slightly less constrained.  We can perform the same calculation to predict the role GCs may have played in the reionization of the Universe by substituting the $\omega_{gc}$ as derived by \cite{Ricotti2002} with the one found empirically within our simulation.  

In the KR13-bis model, 122 GCs formed in the system at $z>7$, and thus, on average, the total mass in GCs during the reionization epoch is $M_{gc}(z>7)=N_{gc}(z>7)\langle m_{gc}\rangle^{ini} = 6.34 \times 10^7$~M$_\odot$.  Thus, $\omega_{gc}f_{ri}=M_{gc}(z>7)/(M_{MW}\Omega_{b}/\Omega_{dm})=4.11\times10^{-4}$ and we determine $M_{MW}$ using Equation 18 with the maximum circular velocity of the main halo.  Using this value along with $\Omega_{DM}/\Omega_b$ from \cite{Planck}, we determine that $N_{ph}^{gc}=4.04$.  Despite this more conservative value, this calculation also yields a value of $N_{ph}^{gc}$ that is in reasonable agreement with the previous calculation.

The number of ionizing photons per baryon per Hubble time $N_{ph}^{gc}$ needed to reionize and maintain the ionization of the IGM at redshift $z$ is $N^{gc}_{ph}=1+t_H/t_{rec}$, where $t_{rec}$ is the hydrogen recombination time. If we define a clumping factor of the IGM $C \equiv \langle n^2\rangle/\langle n\rangle^2$, we find 
\begin{equation}
t_H/t_{rec} \approx 0.68 C \left(\frac{1+z}{8}\right)^{1.5}.
\end{equation}
Thus, assuming $C=2.11$ as found in recent simulations \citep{Shull2012}, we need $N_{ph}^{gc} = 2.43$ ionizing photons escaping galaxies to keep the IGM ionized at $z=7$.  Our estimate of $N_{ph}^{gc}=6.7$ from GCs at $z=7$ implies that if $f_{esc}>36\%$, radiation from GCs alone is sufficient for reionization. We also point out that our estimate of $N_{ph}^{gc}$ is rather conservative because we have assumed $M_{gc}^{min}=10^5$~M$_\odot$, a Kroupa IMF and have neglected infant mortality of clusters. For instance, as discussed in the previous section, $M_{gc}^{min}=10^4$~M$_\odot$ would produce 1.4 times more photons, thus with this assumption, $f_{esc} = 26\%$ would be sufficient for reionization.

The value of $f_{esc}$ for GCs is not well constrained. It depends strongly on the formation model of GCs, in particular, whether or not they form deeply embedded in much more massive molecular clouds.  However, a large efficiency of conversion of gas into stars is required by models GCs formation to avoid unbinding the cluster as a result of gas loss due to stellar feedback. A high star formation efficiency has the twofold effect of increasing the feedback energy and the number of photons emitted with respect to the number of absorbing neutral atoms in the unused gas. For this reason, and because GCs are typically found in the outskirts of dwarf galaxies where the gas density is expected to be low, values of $f_{esc} \sim 0.5-1$ are not unreasonable. But regardless of its value, $f_{esc}$ for GCs should be larger when compared to other modes of star formation.

\section{Conclusions}
We have self consistently modeled the formation history and dynamical
evolution of the GC population of a Milky Way type galaxy by
constraining the GCIMF and formation efficiencies to match
observations of GC populations in local, isolated dwarf galaxies
\citep{Geo2010}.

We have used the merger tree from the Via Lactea II simulation and
GC orbits are computed in a time varying gravitational potential
after they are either accreted from a satellite halo or formed {\it in
  situ}, within the Milky Way halo.  Stellar evolution, two-body
relaxation, dynamical friction, tidal shocks, and tidal truncation are
calculated for each individual cluster in order to reproduce the
observed kinematics and mass function of the observed Milky Way
GC population.

We find that the Galactocentric distances of GCs in our simulations
are very sensitive to the formation efficiencies of GCs
as a function of redshift and halo mass.  Our most accurate model
reproduces the Galactocentric positions, velocities, mass function,
metallicity, and age distributions of GCs in the Milky Way, while being
consistent with the specific frequency $S_N$ of GCs in isolated dwarf
galaxies.  This model predicts that $\sim 38\%$ of the surviving GCs
were formed {\it in situ} while the other $\sim 62\%$ were accreted
from about 20 satellite dwarf galaxies with $v_{cir}>30$~km/s. 

Since we have not tested all possible models for the formation efficiency as a function of redshift and halo mass as well as any other parameter one might conclude the formation efficiency may depend on, we cannot say that the double peaked model is the only way to reproduce the observations in the Milky Way.  However, this model provides a natural explanation for the observable properties of the Milky Way and local dwarf GC populations while also conforming to the constraints on when GCs can form as outlined by \cite{Katz2013}.  For these reasons, our model provides a very likely scenario for the formation and evolution of GCs in hierarchical cosmology.

Our most accurate models reveals two distinct peaks in the GC
formation efficiency at $z \sim 2$ and $z \sim 7-12$ and a GC
formation efficiency that is either remains constant or increases with
decreasing halo mass, contrary $f_* \equiv M_*/M_h$ in present day
galaxies that instead declines rather steeply with decreasing halo. 

Thus, we expect that GC formation was likely a dominant mode of star
formation at least in a subset of dwarf galaxies at high-redshift
whose remnants in the present day universe can be identified as early
type dwarf galaxies. This trend with halo mass combined with evidence of a
peaking GC's formation efficiency at redshifts $z>6$ found in this
work as well as by \cite{Katz2013}, using a completely independent
method, supports the notion that GCs may have played a dominant role
in the reionization of the intergalactic medium \citep{Ricotti2002}.

\subsection*{ACKNOWLEDGMENTS}
We thank the anonymous referee for revising and improving the manuscript.  We also acknowledge Nate Bastian for his very helpful comments and suggestions.  HK's research is supported by Foundation Boustany, Cambridge
Overseas Trust, and the Isaac Newton Studentship.  MR's research is
supported by NASA grant NNX10AH10G and NSF CMMI1125285. This work made
in the ILP LABEX (under reference ANR-10-LABX-63) was supported by
French state funds managed by the ANR within the Investissements
d'Avenir programme under reference ANR-11-IDEX-0004-02.

%.......................................................................
\bibliographystyle{./mn2e}
\bibliography{./HKMR2014.bib}

\appendix
\section{Relaxing the Approximation of Constant Initial Density of GCs}

\begin{figure*}
\centerline{\epsfig{figure=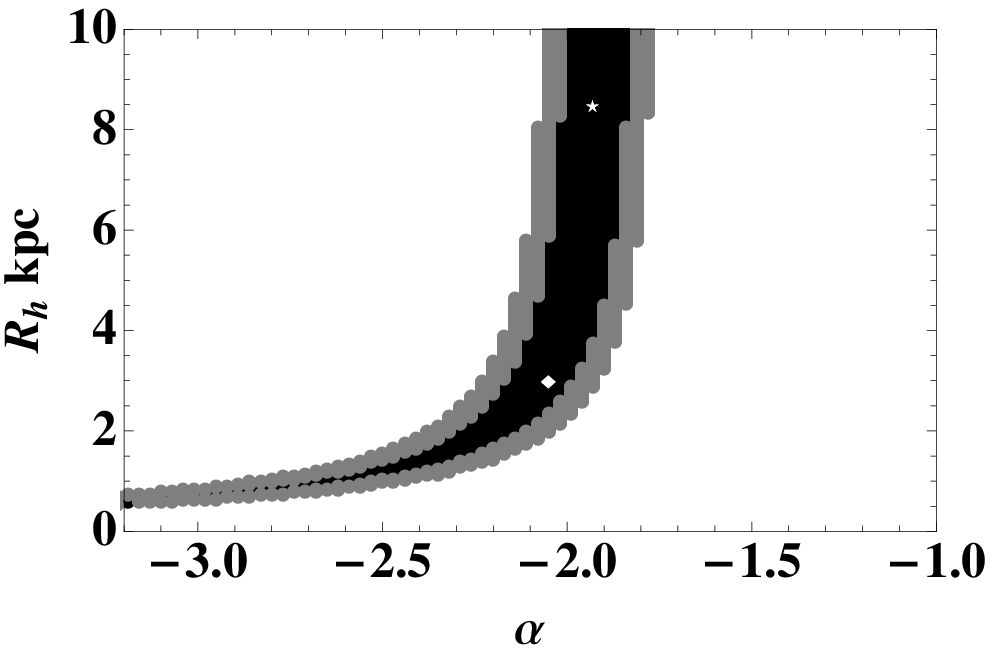,scale=0.87}\epsfig{figure=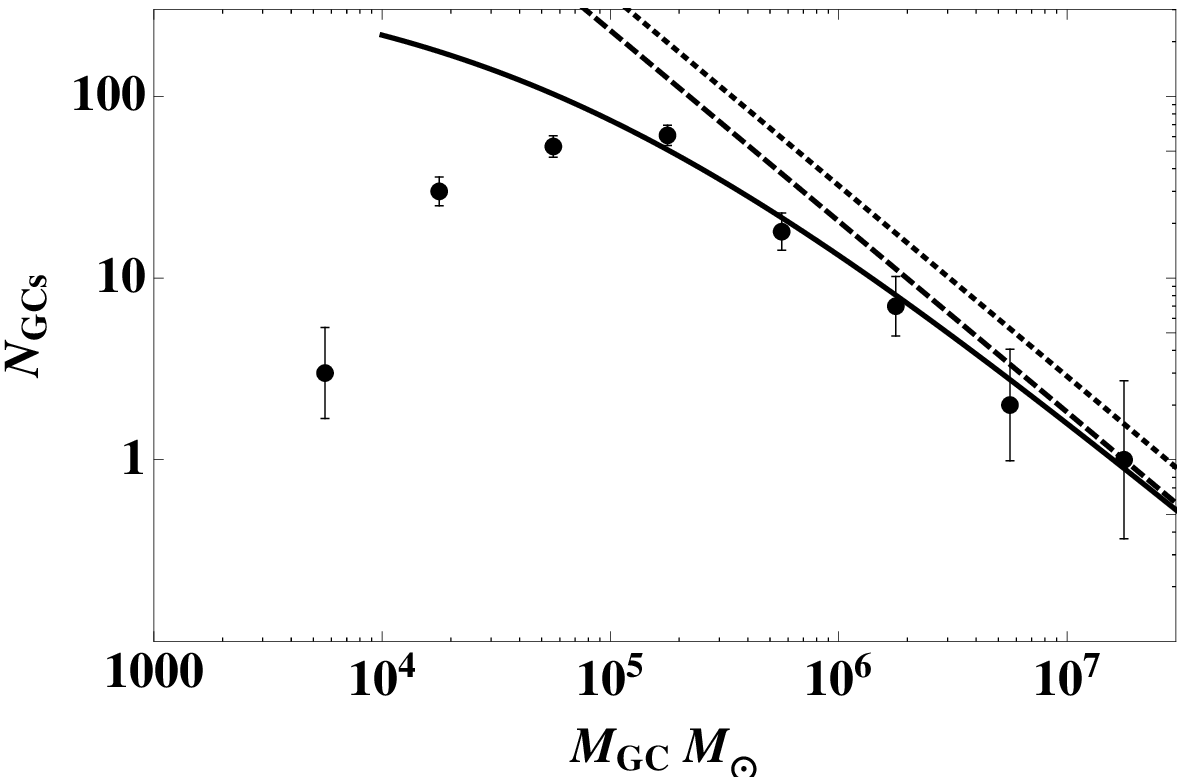,scale=0.7}}
\caption{{\it Left.} Parameter space analysis of the best fit slope
  and initial half mass radius for the GCIMF for the UIRM.  The star
  is the best fit parameters and the black and gray regions are the
  $1\sigma$ and $2\sigma$ confidence levels respectively.  The diamond
  represents the actual parameters we have adopted so the initial
  $R_h$ is much closer to that of the Milky Way GCs. {\it Right.} Our
  chosen GCIMF parameters compared with the dwarf galaxy GCMF for the
  UIRM.  Data points represent the local dwarf galaxy GCMF from
  \protect\cite{Geo2009,Geo2009a}.  The dotted line is the GCIMF prior to stellar
  evolution.  The dashed black like is the GCMF after stellar
  evolution and the solid black line is the GCMF after undergoing
  two-body relaxation for 12~Gyr}
\label{minx3}
\end{figure*}

We have assumed that all GCs, regardless of mass have constant
density (hereafter, we refer to this model as UD [universal density]
model).  We keep the assumption that GCs maintain a constant density as
they evolve, but we explore the case in which the initial density of
each GC is related to their mass.  To test this idea, we assume that
all GCs have a constant $R_h$ so that $\rho_h \propto M_{gc}$
(universal initial radius [UIR] model).  We run a similar parameter
space exploration as was done for the UD model by leaving the initial
$R_h$ and $\alpha$ as free parameters; however we only compare to the
high mass end of the local dwarf galaxy GCMF where tidal effects are
negligible.  The left panel of Figure~\ref{minx3} shows the results of
this simulation.  The current average $R_h$ for Milky Way GCs is
$2.4$~pc.  Once again assuming that $30\%$ of the mass is initially
lost due to stellar evolution, if the GCs remain at the same density,
the minimum necessary $R_h$ is $\sim 2.7$ pc.  The initial $R_h$ is
likely larger since there are other effects that remove mass from the
GC.  If we choose the same slope of $\alpha=-2.05$ for the GCIMF and
set $R_h=3$~pc at formation and evolve this GCIMF via stellar
evolution and two-body relaxation, we find that the resulting GCMF is
no longer consistent with the mass function of local dwarf GCs (see
the right panel of Figure~\ref{minx3}).  The data after the peak fits
as well as when we assumed that all GCs had the same density
regardless of mass.  Tidal effects may begin to play a role towards
the lower end of the mass spectrum which will likely change the shape.
In this scenario, a $10^4$~M$_{\odot}$ GC will have
$\rho_h=88.4$~M$_{\odot}/$pc$^3$, which is over an order of magnitude
lower than the previous model.

The dwarf galaxy GC mass function can also reveal the maximum mass GC
that can form.  In this sample, the maximum mass GC has a mass of
$1.07 \times10^7$~M$_{\odot}$ and thus our GCIMF is require to produce
GCs of at least this mass.  We choose a maximum mass of
$2.9 \times10^7$~M$_{\odot}$, but note that the probability of
producing an object of this size is minimal.  Our simulations are much
less sensitive to the upper bound of the mass function compared to the
lower bound so any reasonable deviation to the upper bound will
produce similar results.

For the UIR model we find an average KS probability of $2.7\%$ with a
maximum probability of $3.8\%$ which is clearly lower than what was
found for the UD model.  Either the UD model better represents the actual
initial properties of GCs in dwarf galaxies or the tidal effects which
we have not included are very important.  \cite{PG08} have run models
similar to the UIRM and found that the resulting mass function is
inconsistent with that of the Milky Way.

While it remains difficult to prove the exact shape of the GCIMF,
whether it be a power law, Gaussian, or some other function, all
shapes will look very similar towards the high mass end and have a slope
approaching $\alpha=-2$.  It may be reasonable to look to the open
cluster mass function and compare with GC.  Surprisingly, the high
mass open clusters also have a power law IMF with a slope of
$\alpha=-2$ which is reasonably consistent with what we have found for
GCs \citep{Zinnecker2009}.

\section{Minimum Formation Efficiencies}
\label{app:form}

Implementing the the minimum formation efficiencies derived in Section
\ref{subsec:constraints}, for both red and blue galaxies, into our
simulations is not straight forward, due to our inability to
differentiate between these two types of galaxies in our simulation.
For this reason, we take an alternative approach to determine an
alternative set of formation efficiencies, independent of the type of
galaxy.  We will find that we can accurately reproduce the mean number
of galaxies that should contain at least one GC as a function of
absolute magnitude, consistent with the observations of
\cite{Geo2010}.

We can define the average mass of a GC from our GCIMF as follows:
\begin{equation}
m_{avg}=\frac{\int^{M_{up}}_{M_{low}}M'\times M'^{\alpha}dM'}{ \int^{M_{up}}_{M_{low}}M'^{\alpha}dM'}.
\end{equation}
The cumulative distribution function (CDF) of the GCIMF can be calculated by:
\begin{equation}
CDF(M)=\frac{\int^{M}_{M_{low}}M'^{\alpha}dM'}{ \int^{M_{up}}_{M_{low}}M'^{\alpha}dM'}.
\end{equation}
If we integrate the derivative of the CDF from the minimum mass needed to survive two-body relaxation and stellar evolution, $M_{surv}$, up to the upper bound of the GCIMF, we can find the probability of drawing a GC from the GCIMF that will survive a specific amount of time.  The probability of survival is defined as:
\begin{equation}
P(M_{surv})=\int^{M_{up}}_{M_{surv}}\frac{\partial CDF(M)}{\partial M}dM.
\end{equation}
Since their are roughly 50 galaxies in Figure~4 in \cite{Geo2010} with absolute visual magnitudes brighter than $-16.5$, the magnitude at which nearly all galaxies brighter have at least one observed GC, the probability that randomly sampling from our GCIMF results in a galaxy of this brightness having no GCs with a mass greater than $M_{surv}$ must be less than $\sim1/50$ or $\sim2\%$.  The number of GCs needed to assure that a galaxy has at least one GC with a mass greater than $M_{surv}$ is then:
\begin{equation}
N_{GCs}=\lceil\frac{\log(1/50)}{\log(1-P(M_{surv}))}\rceil
\end{equation}
where $\lceil x \rceil$ is the ceiling function.  Thus if we assign $N_{GCs}$ to a galaxy, there is only a $\sim2\%$ probability that stellar evolution and two-body relaxation alone will destroy all the GCs in that galaxy.

The specific formation efficiency of old GCs, $\eta$, describes the relation between the the mass of the host halo and the mass of GCs and is heavily constrained by observations.  Assuming that the mass in GCs is proportional to the mass of the host halo:
\begin{equation}
M_{GCs}=\eta M_h
\end{equation}

Using Equations~(13) and (14) from \cite{Geo2010}, and the halo mass of $M_{h,-16.5}$ (which corresponds to an $M_V=-16.5$) such that all galaxies with mass greater than $M_{h,-16.5}$, have at least one GC, we can calculate $\eta$ directly by knowing a value of $N_{GCs}$ as follows:
\begin{equation}
\eta\approx\frac{P(M_{surv})N_{GCs}(0.7m_{surv}-m_{2br})}{M_{h,fid}}
\end{equation}
where $m_{2br}$ is the mass loss due to two-body relaxation, the 0.7 is a result of assuming 30\% of the mass is initially lost to stellar evolution, and $m_{surv}$ is the average mass of a GC in the GCIMF with masses greater than $M_{surv}$.
\begin{equation}
m_{surv}=\frac{\int^{M_{up}}_{M_{surv}}M'\times M'^{\alpha}dM'}{ \int^{M_{up}}_{M_{surv}}M'^{\alpha}dM'}
\end{equation}
The value of $\eta_i$ can also be calculated by:
\begin{equation}
\eta_i=\frac{N_{GCs}m_{avg}}{M_{h,fid}}
\end{equation}

\begin{figure}
\epsfig{figure=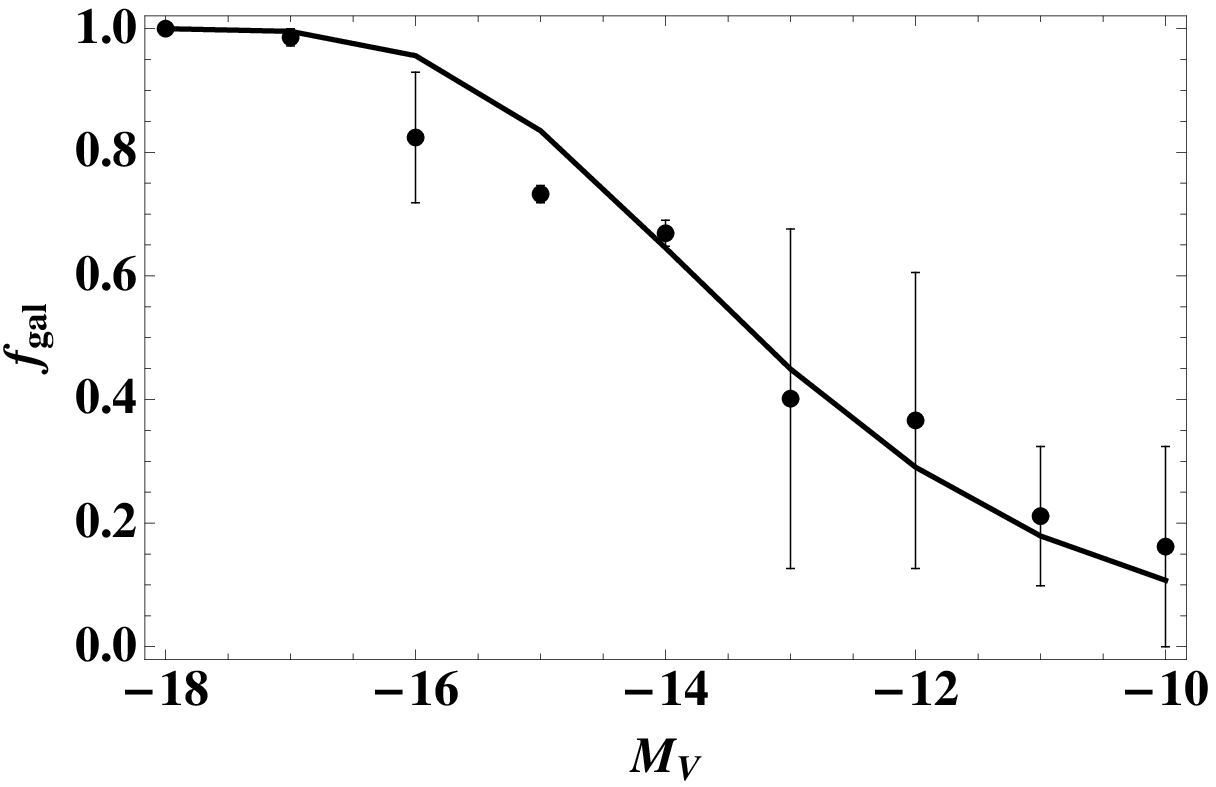,scale=0.65}
\caption{The percentage of galaxies expected to host at least one GC as a function of the absolute visual magnitude of the host galaxy.  The solid line is the expectation using the efficiency for GC mass loss after 12 Gyr and the data points are the average of the blue and red galaxies that host at least one GC from \protect\cite{Geo2010}.  Error bars on data points correspond to the range between red and blue galaxies and the true dispersion in the mean of all galaxies is likely larger than what is plotted.}
\label{fgal_app}
\end{figure}

This derivation is clearly independent of the type of galaxy and we can see, in Figure \ref{fgal_app}, that we have once again reproduced the mean number of galaxies expected to host at least one GC as a function of absolute visual magnitude.

%\bsp % ``This paper has been produced using the ...''
\label{lastpage}
\end{document}